\documentclass[12pt]{article}

\usepackage{newtxtext,newtxmath}

\usepackage{graphicx}
\usepackage{siunitx}
\usepackage{booktabs}
\usepackage{makecell}
\usepackage[letterpaper,margin=1in]{geometry}

\renewenvironment{abstract}
	{\quotation}
	{\endquotation}

\date{}

\makeatletter
\renewcommand{\fnum@figure}{\textbf{Figure \thefigure}}
\renewcommand{\fnum@table}{\textbf{Table \thetable}}
\makeatother

\usepackage{scicite}

\usepackage{url}
\usepackage{xspace}

\newcommand{\Rj}{\ensuremath{R_{\rm{Jup}}}\xspace}
\newcommand{\Mj}{\ensuremath{M_{\rm{Jup}}}\xspace} 
\newcommand{\kms}{km~s$^{-1}$\xspace}

\newcommand{\Lsun}{L_\odot}
\newcommand{\Teff}{\ensuremath{T_\mathrm{eff}}\xspace}
\newcommand{\logg}{\ensuremath{\log{g}}\xspace}
\newcommand{\lbollsun}{\ensuremath{\log(L_\mathrm{bol}/\Lsun)}}

\def\scititle{
	The deuterium-rich atmosphere of a mature super-Jupiter
}
\title{\bfseries \boldmath \scititle}

\author{
Jerry~W.~Xuan$^{1,2\ast}$,
Yayaati~Chachan$^{3}$,
Jean-Baptiste~Ruffio$^{4}$,
Kazumasa~Ohno$^{5}$,\and
Shang-Min~Tsai$^{7}$,
James~J.~Mang$^{6}$,
Yapeng~Zhang$^{9}$,
Aniket~Sanghi$^{9}$,\and
Bj\"orn~Benneke$^{2}$,
Alexis~Bidot$^{8}$,
Geoffrey~A.~Blake$^{11}$,
Brendan~P.~Bowler$^{1}$,\and
Mathilde~M\^alin$^{8,10}$,
Elisabeth~Matthews$^{14}$,
Caroline~V.~Morley$^{6}$,
Ruth~Murray-Clay$^{3}$,\and
Marshall~Perrin$^{8}$,
Jason~J.~Wang$^{17,18}$,
Carrie~He$^{2}$,
Katelyn~Horstman$^{9}$,\and
Heather~A.~Knutson$^{11}$,
Alexander~Madurowicz$^{8}$,
Christian~Marois$^{12,13}$,\and
Dimitri~Mawet$^{9,15}$,
Rebecca~Oppenheimer$^{16}$,
William~Thompson$^{12}$\and
\small$^{1}$Department of Physics, University of California, Santa Barbara, CA 93106, USA.\and
\small$^{2}$Department of Earth, Planetary, and Space Sciences,
University of California, Los Angeles, CA 90095, USA.\and
\small$^{3}$Department of Astronomy \& Astrophysics,
University of California, Santa Cruz, CA 95064, USA.\and
\small$^{4}$Department of Astronomy \& Astrophysics,
University of California, San Diego, La Jolla, CA 92093, USA.\and
\small$^{5}$Division of Science, National Astronomical Observatory of Japan,
Mitaka, Tokyo 181-8588, Japan.\and
\small$^{6}$Department of Astronomy, The University of Texas at Austin,
Austin, TX 78712, USA.\and
\small$^{7}$Institute of Astronomy and Astrophysics, Academia Sinica,
Taipei 10617, Taiwan.\and
\small$^{8}$Space Telescope Science Institute,
Baltimore, MD 21218, USA.\and
\small$^{9}$Department of Astronomy, California Institute of Technology,
Pasadena, CA 91125, USA.\and
\small$^{10}$Department of Physics \& Astronomy, Johns Hopkins University,
Baltimore, MD 21218, USA.\and
\small$^{11}$Division of Geological \& Planetary Sciences,
California Institute of Technology, Pasadena, CA 91125, USA.\and
\small$^{12}$NRC Herzberg Astronomy and Astrophysics,
Victoria, BC V9E 2E7, Canada.\and
\small$^{13}$Department of Physics and Astronomy, University of Victoria,
Victoria, BC V8P 5C2, Canada.\and
\small$^{14}$Max-Planck-Institut f\"ur Astronomie,
K\"onigstuhl 17, 69117 Heidelberg, Germany.\and
\small$^{15}$Jet Propulsion Laboratory, California Institute of Technology,
Pasadena, CA 91109, USA.\and
\small$^{16}$Department of Astrophysics, American Museum of Natural History,
New York, NY 10024, USA.\and
\small$^{17}$Department of Physics and Astronomy, Northwestern University,
Evanston, IL 60208, USA.\and
\small\parbox{0.92\textwidth}{\centering
$^{18}$Center for Interdisciplinary Exploration and Research in Astrophysics,
Northwestern University, Evanston, IL 60201, USA.}\and
\small$^\ast$Corresponding author. Email: jxuan@ucsb.edu
}

\begin{document} 

\maketitle

\begin{abstract} \bfseries \boldmath
The deuterium-to-hydrogen ratio (D/H) has provided key insights into the origin and evolution of the Solar System. Using the James Webb Space Telescope, we detected deuterated methane at $48\sigma$ in a mature giant exoplanet, $\epsilon$~Ind~Ab. We measure the planet's D/H ratio to be $(4.4\pm0.7)\times10^{-5}$, which is similar to those of Uranus and Neptune but about twice those of Jupiter, Saturn, and the protosolar nebula. $\epsilon$~Ind~Ab's heavy elements (C, O, N, and S) are uniformly enriched to a level similar to Jupiter's. These measurements require $\epsilon$~Ind~Ab to have accreted solids with $\rm D/H\gtrsim1\%$, at least 30 times more deuterium-rich than typical Solar System comets. Such D/H values are found in the interstellar comet 3I/ATLAS and in organic molecules in star-forming regions, indicating that $\epsilon$~Ind~Ab accreted pristine solids formed in a very cold environment. 
\end{abstract}

\noindent
Deuterium is a heavy isotope of hydrogen with one proton and one neutron in its nucleus. It was formed in the Big Bang with a primordial abundance relative to hydrogen of $(2.527\pm 0.030)\times10^{-5}$ \cite{Cooke2018}. Since then, deuterium has been gradually destroyed by burning in stellar interiors, and the local interstellar medium (ISM) today has $\rm D/H\geq(2.0\pm0.1)\times10^{-5}$ \cite{Prodanovic2010}. In cold star-forming regions, chemical fractionation redistributes deuterium among molecules and can raise the D/H of certain ices and organics by orders of magnitude above the bulk abundance \cite{Ceccarelli2014}. In the Solar System, D/H has been used to study the formation conditions of comets \cite{BockeleeMorvan2015, Muller2022}, the origin of Earth's ocean water \cite{Alexander2012, Cleeves2014}, and the loss of water from Venus and Mars due to atmospheric escape \cite{Donahue1982, Villanueva2015}. Jupiter and Saturn have D/H close to the protosolar value of $\sim2\times10^{-5}$ \cite{Lellouch2001, Pierel2017}, whereas Uranus and Neptune are enriched in deuterium by a factor of two, usually attributed to the accretion of icy planetesimals, whose remnants survive as comets \cite{Hubbard1980, Feuchtgruber2013}. For giant planets, D/H could therefore constrain the nature of the solids they accreted during formation.

$\epsilon$~Ind~Ab is the first mature ($\approx4$ Gyr) cold Jupiter imaged around another star \cite{Matthews2024}. It has a dynamical mass of $6-8$ Jupiter masses ($\Mj$) \cite{Sanghi2026, Matthews2026} and orbits with a semi-major axis of $16\pm2$ astronomical units (AU) around a nearby K dwarf ($3.6$ parsec). Given its cold temperature of $\Teff\approx 275~$K, $\epsilon$ Ind Ab opens new frontiers in exoplanet atmospheric studies. For example, deuterated methane is predicted to be most readily detectable in cold exoplanets and brown dwarfs with $\Teff\lesssim600~$K \cite{Morley2019}, and was recently detected in the field brown dwarf WISE 0855 which has an effective temperature similar to that of $\epsilon$~Ind~Ab \cite{Rowland2024}. 

\begin{figure}[h]
\centering
\includegraphics[width=\linewidth]{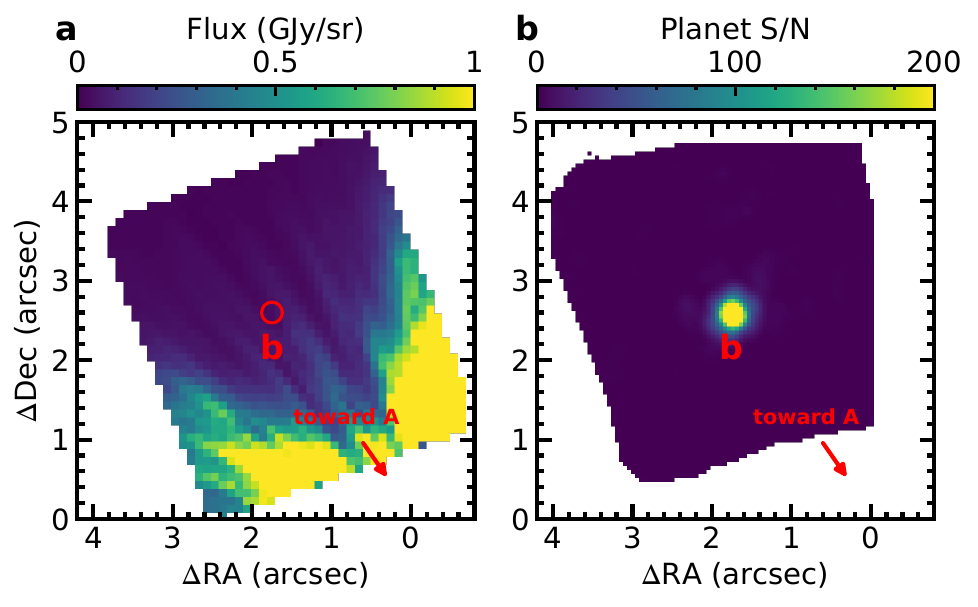}
\caption{\textbf{Detection of $\epsilon$ Ind Ab with moderate-resolution mode of JWST/NIRSpec IFU in the $3-5~\mu$m spectral range.} \textbf{a,} The median spectral cube for one of the two NIRSpec rolls from the standard JWST pipeline before starlight subtraction. Here, the stellar diffraction pattern dominates the image. The host star $\epsilon$ Ind A is just outside the field of view of the IFU, and the red arrow marks its direction. \textbf{b,} S/N map for $\epsilon$ Ind Ab combining the two NIRSpec rolls and both NIRSpec detectors. Summed over all wavelength channels, the planet is detected with a S/N of approximately 600. 
}
\label{fig:detection}
\end{figure}

We present moderate-resolution spectroscopy of $\epsilon$ Ind Ab observed with the integral field unit (IFU; $R\approx2700$, $2.85-5.30~\mu$m) of JWST's near-infrared spectrograph (NIRSpec). The observations were obtained on 25 October 2025 UT as part of the cycle 4 program 8714 (PI J. Xuan) and had a total integration time of 4.4 hours. The observations were equally split into two sequences at different observatory roll angles, and each roll includes data on two detectors, NRS1 and NRS2. The planet was placed at the center of the IFU's field-of-view (FoV; $3^{\prime\prime}$-by-$3^{\prime\prime}$). At the time of observations, $\epsilon$ Ind Ab's projected separation from the host star was $3.55^{\prime\prime}$, so the core of the stellar point-spread function (PSF) is outside the FoV. Nevertheless, because the star is nearly 1 million times brighter than the planet at $4.1~\mu$m \cite{Sanghi2026}, diffraction features from the stellar PSF dominate the raw data cube (Fig.~\ref{fig:detection}). Due to the large projected separation, however, the local stellar halo is only about 2 times brighter than the planet in NRS2 ($4.1-5.3~\mu$m). 

\begin{figure}[h]
\centering
\includegraphics[width=\linewidth]{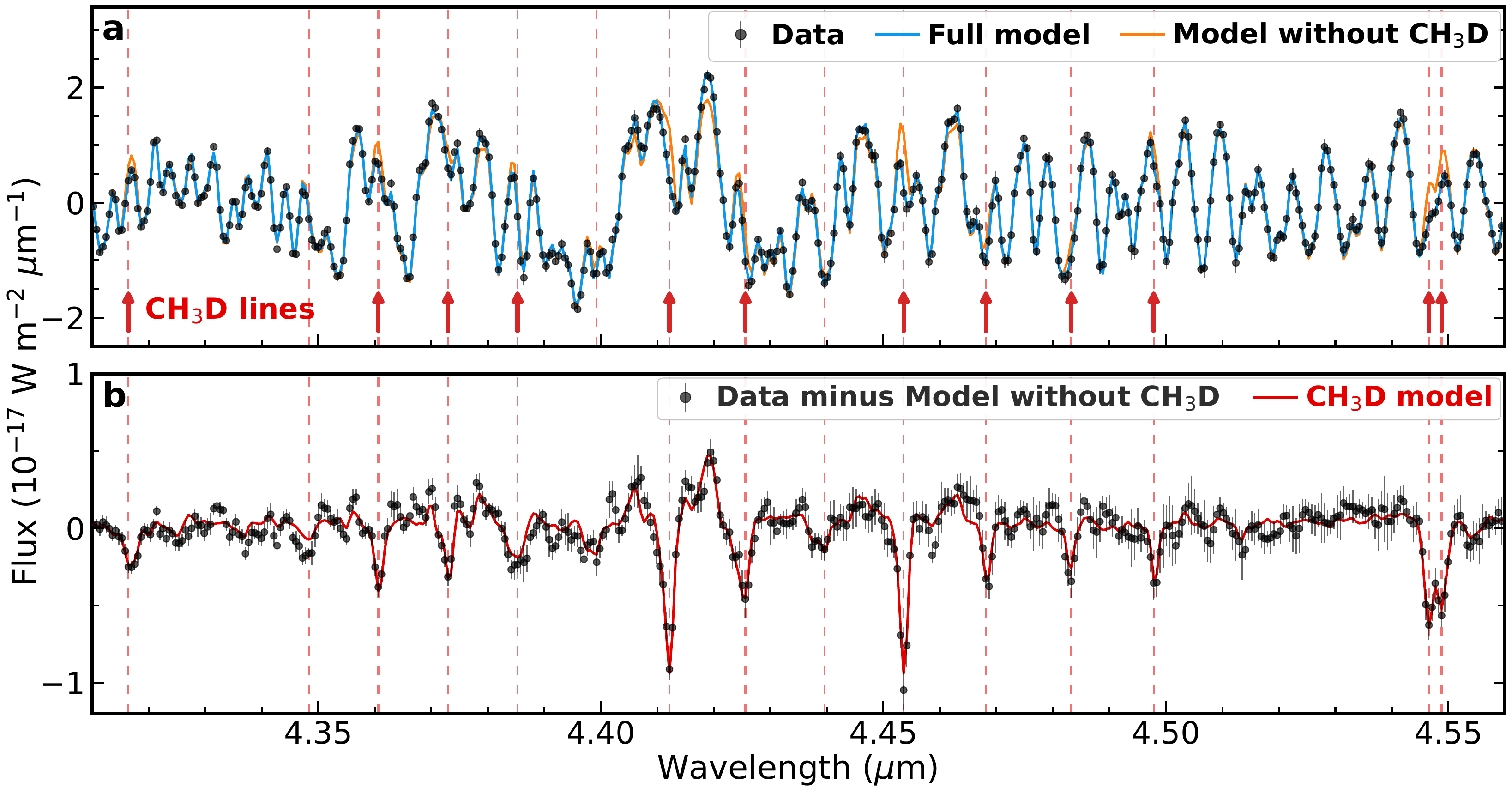}
\caption{\textbf{Detection of CH$_3$D in $\epsilon$ Ind Ab.} \textbf{a,} A segment of the NIRSpec spectrum (black). The best-fit full model including all species is in blue, and the best-fit model from a separate retrieval that excludes CH$_3$D opacity is in orange. The difference between the blue and orange models is due to CH$_3$D; we mark locations of CH$_3$D lines with red dashed lines and arrows. \textbf{b,} The residuals after subtracting the best-fit model without CH$_3$D from the data (black), plotted against the CH$_3$D-only model (red). A cross-correlation analysis between the residuals and CH$_3$D model indicates a detection $\rm S/N\approx48$ for CH$_3$D in $\epsilon$ Ind Ab's atmosphere. Equivalent plots to panel b for other species are shown in figure~\ref{fig:other_detections}.
}
\label{fig:spectrum}
\end{figure}

To detect $\epsilon$ Ind Ab and extract its spectrum, we take advantage of the spectral difference between the $\Teff\approx275~$K planet and $\Teff\approx4700~$K star, which show vastly different absorption lines. Specifically, we use the \texttt{breads} package \cite{Ruffio2024} which is frequently used in similar high-contrast observations from JWST/NIRSpec \cite{Madurowicz2025, Ruffio2026, Xuan2026, Baburaj2026, 2026arXiv260911827H, 2026arXiv260911828M}. To detect $\epsilon$ Ind Ab, we simultaneously model the planet and the starlight directly in the NIRSpec detector frame to avoid systematic extraction issues from the spatial undersampling of the NIRSpec IFU (see Methods). To extract its spectrum, we first subtract the starlight row-by-row on the detector using a spline-based model, and then fit a model PSF to the combined dither-sequence at the planet's position in the field of view for each wavelength \cite{Ruffio2026}. This provides a high-pass filtered spectrum of $\epsilon$ Ind Ab, in which the spectral features of the star have been removed. The detection map of $\epsilon$ Ind Ab is shown in Fig.~\ref{fig:detection}, while the spectrum is shown in Fig.~\ref{fig:spectrum} and figure~\ref{fig:nirspec_and_phot}.

To measure the elemental and isotopic abundances of $\epsilon$ Ind Ab, we carried out Bayesian retrieval analyses using the radiative transfer code \texttt{petitRADTRANS} \cite{molliere_petitRADTRANS_2019, Nasedkin2024_PRT}. In addition to the NIRSpec spectrum, we also jointly model $\epsilon$ Ind Ab's $4-25\mu$m photometry \cite{Matthews2024, Matthews2026, Sanghi2026}, which anchors the broadband continuum shape (see figure~\ref{fig:nirspec_and_phot}). The two NIRSpec rolls are fitted separately, which allows us to compare the consistency between two independent datasets. The parameters of the atmospheric model include the planet's mass, radius, radial velocity (RV) shift, and those describing the thermal structure, chemical abundances, and clouds in the atmosphere (see Methods). We considered both free retrievals, where the mass-mixing ratios of each molecule are fitted independently and constant with pressure, as well as chemically-consistent retrievals \cite{Zhang2023_AFLep, Ruffio2026}, where the individual molecular abundances are determined by four elemental abundance parameters [C/H], [O/H], [S/H], and [N/H] and an equilibrium chemistry network. In the chemically-consistent retrievals, disequilibrium chemistry is accounted for by fitting the vertical eddy diffusion coefficient, $K_{\rm zz}$, using the quench approximation\cite{Zahnle_methane_2014} (see Methods). To account for possible underestimation of the data uncertainties, we fit multiplicative error inflation factors \cite{Xuan2022}. NRS1 requires no inflation whereas the NRS2 uncertainties need to be inflated by a factor of 2 to yield residuals with a reduced $\chi^2$ of unity (Fig.~\ref{fig:spectrum}).

We detect twelve different molecular isotopologues -- $^{12}$CH$_4$, $^{12}$CH$_3$D, $^{13}$CH$_4$, H$_2^{16}$O, H$_2^{18}$O, H$_2^{17}$O, $^{12}$CO, $^{13}$CO, C$^{18}$O, CO$_2$, H$_2$S, and NH$_3$ -- in the atmosphere of $\epsilon$ Ind Ab (Fig.~\ref{fig:spectrum} and figure~\ref{fig:other_detections}). The molecular detections were validated using leave-one-out tests and cross-correlation analyses \cite{Zhang2021Natur.595..370Z, Xuan2026, methods}. The elemental abundances of [C/H], [O/H], [S/H], and [N/H], as measured by the chemically-consistent retrievals, are consistent between the two NIRSpec rolls within $0.1$ dex. The statistical error on the abundances can be as small as $0.02$ dex. Therefore, for all planet parameters reported here, we quote the mean of the posterior medians from the two rolls, and add the scatter between rolls in quadrature as a systematic error. From the chemically-consistent retrievals, we measure ${\rm C/H}=5.7^{+1.0}_{-0.9}$, ${\rm O/H}=5.2^{+1.1}_{-0.9}$, ${\rm S/H}=5.3^{+2.7}_{-1.8}$, and ${\rm N/H}=4.5^{+1.4}_{-1.1}$ times the stellar values  \cite{Baburaj2026b}. The free retrievals yield consistent elemental abundances \cite{methods}. Therefore, $\epsilon$ Ind Ab exhibits a nearly uniform heavy element enrichment pattern across volatile species (C, O, N) and refractory species (S). The same pattern is seen in Jupiter\cite{Guillot2023} and systems of young super-Jupiters \cite{Ruffio2026, Xuan2026b}.

From the retrieved abundances of $^{12}$CH$_4$ and $^{12}$CH$_3$D, we obtain $\rm (D/H)_{CH_4} = (5.4\pm0.2) \times10^{-5}$. Since the atmosphere of a gas giant like $\epsilon$ Ind Ab is mainly composed of H$_2$, the D/H from H$_2$ and HD is a better tracer of the overall D/H. However, H$_2$ and HD are not directly detectable in our NIRSpec spectrum. Thus, a correction factor $f=\rm (D/H)_{CH_4} / \rm (D/H)_{H_2}$ is required to account for the difference between the D/H measured from CH$_4$ and D/H in hydrogen \cite{Lecluse1996}. From both analytic and numerical calculations \cite{methods}, we estimate that $f = 1.10$--$1.35$ for $\epsilon$ Ind Ab, which yields a corrected $\rm (D/H)_{H_2}=(4.4\pm0.7)\times10^{-5}$ for the planet's atmosphere. We note that our inferred $f$ values are similar to those found for Jupiter and Saturn, which are $f\approx1.25$ and $f\approx1.38$, respectively \cite{Lecluse1996}. 

\begin{figure}[h]
\centering
\includegraphics[width=\linewidth]{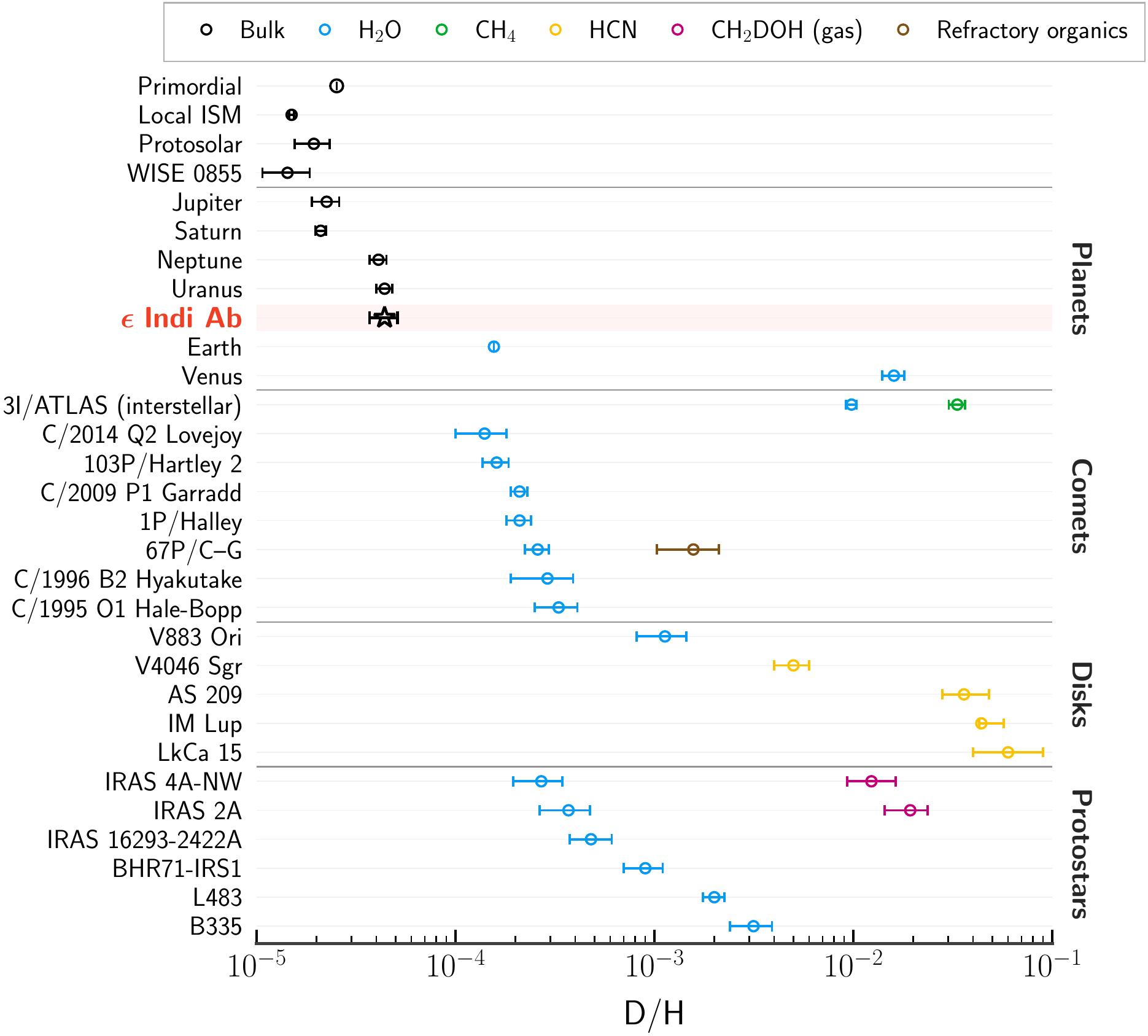}
\caption{\textbf{D/H measurements in various astrophysical environments, including planets, comets, circumstellar disks, and protostars.} The black markers indicate estimates of bulk D/H, either measured directly from atomic or molecular hydrogen or inferred from other molecules with appropriate corrections (CH$_3$D/CH$_4$ for $\epsilon$ Ind Ab and WISE 0855). The other colors denote D/H in specific molecular tracers, and do not represent bulk D/H: H$_2$O (blue), CH$_4$ (teal), HCN (yellow), gas-phase deuterated methanol, CH$_2$DOH (magenta), and refractory organics (brown). Data sources: primordial value~\cite{Cooke2018}; WISE~0855~\cite{Rowland2024}; local ISM, Solar System planets, most Solar System comets, and protostars, from a published compilation~\cite{Nomura2023}; 3I/ATLAS~\cite{Cordiner2026,2026arXiv260320445R}; 67P/Churyumov--Gerasimenko, using updated values~\cite{Mandt2024,Paquette2021}; V883~Ori~\cite{Tobin2023}; HCN in other protoplanetary disks~\cite{Huang2017}.
}
\label{fig:dtoh}
\end{figure}

The D/H of $\epsilon$ Ind Ab is comparable to those of Uranus and Neptune and is approximately twice the protosolar value of $2\times10^{-5}$ (Fig.~\ref{fig:dtoh}). The elevated D/H of the ice giants can be explained by accretion of ice with D/H at or below cometary values \cite{Hubbard1980, Feuchtgruber2013}, and their atmospheric C/H of $\sim70-100\times$ solar \cite{Guillot2023} is consistent with this model. The similarity in D/H is therefore striking because $\epsilon$ Ind Ab is a gas giant with a mass of $6-8~\Mj$, and C/H and O/H only $\approx5\times$ stellar. Indeed, Jupiter and Saturn have D/H ratios that agree with the protosolar value within $1\sigma$ error bars \cite{Lellouch2001,Pierel2017}. The nearby brown dwarf WISE 0855 also has a D/H close to the protosolar value \cite{Rowland2024}; as validation of our model, we reproduce the published D/H of WISE 0855 \cite{methods}. Below, we discuss how $\epsilon$ Ind Ab may have acquired a high D/H ratio. 

First, we consider if the $\epsilon$ Ind system may have formed in a molecular cloud with higher bulk D/H than the ones which formed the Solar System or WISE 0855. As the host star and binary brown dwarf \cite{McCaughrean2004} in the $\epsilon$ Ind system have long burned their deuterium, we cannot directly test this hypothesis. In the local diffuse ISM, the gas-phase D/H varies from $\approx(0.5-2.2)\times10^{-5}$ along different sightlines \cite{Linsky2006,Prodanovic2010}, with the lower values generally attributed to the depletion of deuterium onto dust grains. After correction, the bulk, undepleted D/H of the local ISM is estimated to be $\rm D/H\geq(2.0\pm0.1)\times10^{-5}$ \cite{Prodanovic2010}. Most importantly, since deuterium is gradually destroyed by nuclear burning after the Big Bang, no natal cloud should have a bulk D/H above the primordial value of $(2.527\pm0.030)\times10^{-5}$ \cite{Cooke2018}. Thus, a high D/H from a molecular cloud cannot explain $\epsilon$ Ind Ab's deuterium enrichment. 

Next, we consider accretion of deuterium-rich solids, the same scenario invoked for Uranus and Neptune. In the Solar System, comets exhibit a range of D/H between $\approx(1.4-5.3)\times10^{-4}$ measured from water \cite{Nomura2023}.
The interstellar comet 3I/ATLAS shows a much higher $\rm D/H\approx10^{-2}$ from water \cite{SalazarManzano2026, Cordiner2026}, and $\rm D/H\approx3\times10^{-2}$ from methane \cite{2026arXiv260320445R}, which supports its formation from a colder and less thermally processed environment than that experienced by the Solar System comets. Around the outbursting protostar V883 Ori, a $\rm D/H\approx1\times10^{-3}$ was measured from H$_2$O ice that sublimated into gas phase in the outer disk at $\sim100$ AU distances \cite{Tobin2023}. This value is a factor of a few higher than the D/H of Solar System comets. Subsequently, a high D$_2$O/H$_2$O ratio comparable to values measured in protostellar envelopes was found for V883 Ori, supporting the inheritance of the water in its disk from molecular clouds \cite{Leemker2025}. Such a conclusion has also been reached for water in the Solar System comets \cite{Cleeves2014, Altwegg2017}. In this context, $\epsilon$ Ind Ab could have acquired its high D/H ratio through accretion of deuterium-rich solids from the cold outer regions of its protoplanetary disk, where pristine material from the molecular cloud phase can be preserved. 

To test this hypothesis, we use the composition model \texttt{accomp} \cite{Chachan2026} to fit the elemental abundances (C/H, O/H, S/H, and N/H) of $\epsilon$ Ind Ab along with its D/H ratio \cite{methods}. The uniform enrichment in C, O, N, and S is most consistent with formation outside the N$_2$ or CO snowlines ($\sim 56$ AU and $\sim36$ AU, respectively). Both these snowlines lie beyond the current semi-major axis at $16\pm2$ AU \cite{Sanghi2026}, which implies $\epsilon$ Ind Ab could have formed farther out and migrated inward. Alternatively, icy pebbles originating beyond these snowlines could have drifted inward to enrich the planet at its current location if they consisted of amorphous ice with trapped volatiles such as CO and N$_2$, a mechanism also proposed for Jupiter \cite{Monga2015, Mousis2019}. Accretion of solids from the cold outer disk also helps explain $\epsilon$ Ind Ab's high D/H, since such material is more likely to have been inherited from the protostellar stage without subsequent thermal processing. Assuming that $\epsilon$ Ind Ab formed beyond the N$_2$ snowline, the measured elemental abundances imply that it accreted $89^{+18}_{-15}~M_\oplus$ of metals through solids. To simultaneously match the metal enrichment and the D/H, the accreted solids must contain either water ice with D/H $\sim 1.5\times 10^{-2}$ or refractory organics with D/H $\sim5\times10^{-2}$ (figure~\ref{fig:DtoH_vs_OtoH_CtoH}). These D/H ratios substantially exceed those of Solar System comets (Fig.~\ref{fig:dtoh}) but are comparable to the D/H of 3I/ATLAS \cite{SalazarManzano2026, Cordiner2026, 2026arXiv260320445R} and the D/H of organic molecules observed in protostellar envelopes \cite{Parise2006, Taquet2019}. The required D/H can be a factor of two lower if accreted solids contained both D-rich water ice and organics (figure~\ref{fig:DtoH_water_organics}).

\begin{figure}[h]
\centering
\includegraphics[width=\linewidth]{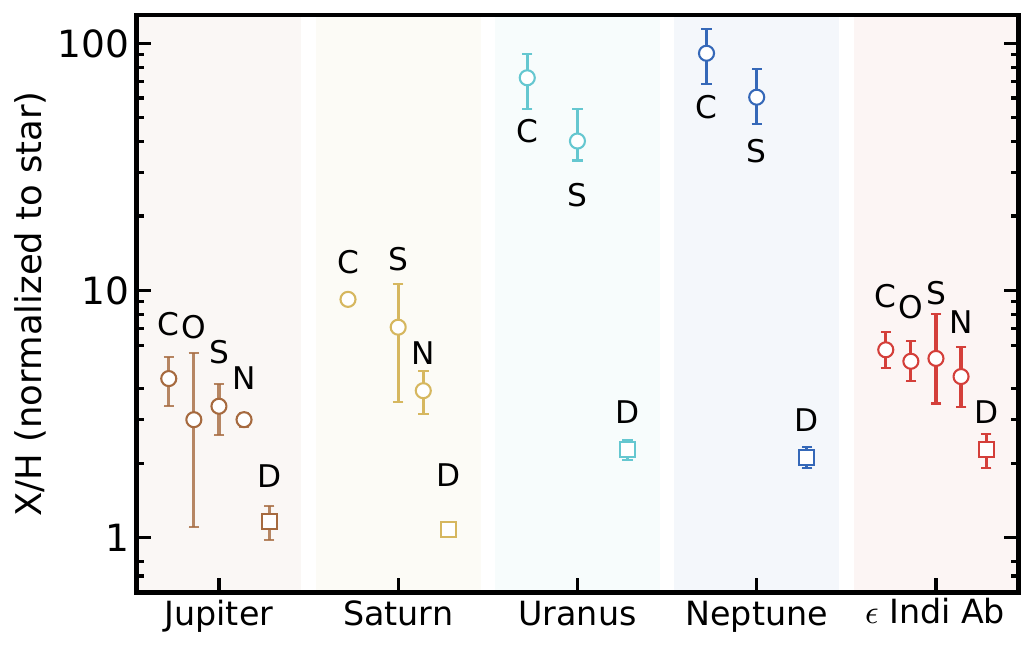}
\caption{\textbf{$\epsilon$ Ind Ab shows similar levels of C, O, S and N enrichment as Jupiter, but a D/H ratio comparable to those of the ice giants.} The abundances of C, O, S, N (circles), and D (squares) relative to H in $\epsilon$ Ind Ab and the Solar System gas and ice giants. All values are normalized to the stellar values: for $\epsilon$ Ind A, we use the stellar abundances \cite{Baburaj2026b}, and for the Solar System planets we normalize to the Sun's photospheric abundances \cite{Asplund2009}. The C, O, S, and N abundances of the Solar System planets are from a published compilation~\cite{Guillot2023}, and their D/H values are from previous measurements~\cite{Lellouch2001,Pierel2017,Feuchtgruber2013}.
}
\label{fig:elements}
\end{figure}

The calculation above assumes that $\epsilon$ Ind Ab is well mixed and that the atmospheric D/H reflects the bulk D/H. Another possibility we consider is late stage accretion of solids combined with inefficient mixing between the upper atmosphere and interior, which could preserve the imprints of any accreted solids in the observed atmosphere. Such a scenario has been proposed to explain Jupiter's atmospheric enrichment \cite{Howard2023, Muller2024}. This scenario could reduce the total \textit{mass} of accreted solids by confining them to a thin atmospheric shell. However, the relative abundance ratios between C, O, S, N, D, and H (Fig.~\ref{fig:elements}) still require the planet to accrete highly deuterium-rich material. Regardless of whether $\epsilon$ Ind Ab is well-mixed or not, the planet's high D/H cannot be explained by accreting Solar System-like comets as doing so would increase the planet's O/H and C/H to much higher values of $\sim200 \times$ stellar, inconsistent with the observed value of $\approx5\times$ stellar (figure~\ref{fig:DtoH_vs_OtoH_CtoH}). Finally, we also consider if disk gas accreted by the planet can become enriched in D and metals during disk dispersal but find that it is implausible \cite{methods}.

As a cold and mature gas giant in the solar neighborhood, $\epsilon$ Ind Ab provides a rare point of comparison with the giant planets of the Solar System. The high D/H of $\epsilon$ Ind Ab suggests that its planetary system inherited pristine ice and refractory organics from the molecular cloud stage, and that these minimally processed solids were accreted by the planet. In the Solar System, deuterium has revealed the inheritance of interstellar ices and organics in meteorites, comets, and Earth's oceans \cite{Busemann2006, Cleeves2014, Muller2022}. Around other stars, protoplanetary disks appear to inherit much of their water ice from the molecular cloud  \cite{Tobin2023, Leemker2025}, and the interstellar comet 3I/ATLAS shows that such pristine material can be preserved as planet building blocks \cite{Cordiner2026, SalazarManzano2026}. $\epsilon$ Ind Ab traces this inheritance to a mature super-Jupiter, demonstrating that interstellar material can be incorporated into planets and leave a sizable imprint even for a gas giant. Future JWST measurements of D/H across a larger sample of cold planets are poised to reveal how planetary systems inherit and process material from molecular clouds, placing the formation of the Solar System in a broader context.


\clearpage 

\bibliography{epsind_nirspec} 
\bibliographystyle{sciencemag}


\section*{Acknowledgments}
J.W.X. thanks Sam de Regt for advice on using \texttt{fastchem} in retrievals, Melanie Rowland for sharing the WISE 0855 spectrum, and Aneesh Baburaj for advice on stellar abundances. We thank Jeehyun Yang, Teresa Paneque-Carreño, Ngoc Tuan Truong, Edward Young, Peng Ni, Jonathan Fortney, Eliza Kempton, and Mark Giovinazzi for helpful discussions. J.W.X. would also like to thank Jingwen Zhang for her support and constructive suggestions.
This work is based on observations made with the NASA/ESA/CSA James Webb Space Telescope. The data were obtained from the Mikulski Archive for Space Telescopes at the Space Telescope Science Institute, which is operated by the Association of Universities for Research in Astronomy, Inc., under NASA contract NAS 5-03127 for JWST. These observations are associated with program 8714.
This work benefited from the 2026 Exoplanet Summer Program in the Other Worlds Laboratory (OWL) at the University of California, Santa Cruz, a program partially supported by funding from NASA.

\paragraph*{Funding:}
J.W.X. acknowledges support from the Heising-Simons Foundation 51 Pegasi b Fellowship (grant \#2025-5887). Y.Z. acknowledges support from the Heising-Simons Foundation 51 Pegasi b Fellowship (grant \#2023-4298).

\paragraph*{Author contributions:}
J.W.X. led the data reduction and atmospheric retrievals, coordinated the analyses, and wrote the manuscript. Y.C. led the composition modeling of the planet and wrote the corresponding section. J-B.R. provided guidance with the data reduction and contributed text. K.O. contributed to the planet formation interpretation and helped with defining the chemical framework used in the retrievals. S-M.T. led the chemical kinetics models with \texttt{VULCAN}. J.M. helped with defining the P-T profile priors. J.W.X., J-B.R., A.S., and Y.Z. designed the GO 8714 observing program. B.B., A.B., G.A.B., M.M., E.M., C.V.M., R.M-C., M.P., J.J.W., B.P.B., H.A.K. provided advice on various aspects of the project. Other authors are part of the GO 8714 team or commented on the manuscript.

\paragraph*{Competing interests:}
There are no competing interests to declare.

\paragraph*{Data and materials availability:}
The JWST/NIRSpec data used in this paper are from JWST GO program 8714 (PI J. Xuan) and are publicly available from the Mikulski Archive for Space Telescopes (MAST)~\cite{MAST_GO8714}. 
The NIRSpec data reduction was performed using the \texttt{breads} package available at \url{https://breads.readthedocs.io/} \cite{breads2024}. The atmospheric models were generated using the \texttt{petitRADTRANS} radiative transfer tool \cite{molliere_petitRADTRANS_2019} available at \url{https://petitradtrans.readthedocs.io/}. 


\subsection*{Supplementary materials}
Materials and Methods\\
Figs. S1 to S7\\
Tables S1 to S3\\
References \textit{(60-\arabic{enumiv})}


\newpage


\renewcommand{\thefigure}{S\arabic{figure}}
\renewcommand{\thetable}{S\arabic{table}}
\renewcommand{\theequation}{S\arabic{equation}}
\renewcommand{\thepage}{S\arabic{page}}
\setcounter{figure}{0}
\setcounter{table}{0}
\setcounter{equation}{0}
\setcounter{page}{1} 


\begin{center}
\section*{Supplementary Materials for\\ \scititle}
Jerry~W.~Xuan$^{\ast}$,
Yayaati~Chachan,
Jean-Baptiste~Ruffio,
Kazumasa~Ohno,
Shang-Min~Tsai,
James~J.~Mang,
Yapeng~Zhang,
Aniket~Sanghi,
Bj\"orn~Benneke,
Alexis~Bidot,
Geoffrey~A.~Blake,
Brendan~P.~Bowler,
Mathilde~M\^alin,
Elisabeth~Matthews,
Caroline~V.~Morley,
Ruth~Murray-Clay,
Marshall~Perrin,
Jason~J.~Wang,
Carrie~He,
Katelyn~Horstman,
Heather~A.~Knutson,
Alexander~Madurowicz,
Christian~Marois,
Dimitri~Mawet,
Rebecca~Oppenheimer,
William~Thompson\\
\small$^\ast$Corresponding author. Email: jxuan@ucsb.edu
\end{center}

\subsubsection*{This PDF file includes:}
Materials and Methods\\
Figures S1 to S7\\
Tables S1 to S3

\newpage


\subsection*{Materials and Methods}

\subsubsection*{Observations}
In this study, we analyze JWST/NIRSpec IFU observations of $\epsilon$ Ind Ab from $2.85-5.30~\mu$m in the medium-resolution spectroscopy mode ($R\sim2,700$; filter: F290LP; grating: G395H) from JWST Cycle 4 GO program 8714 (PI: J. Xuan). The observations took place on 25 October 2025 (UT). 
The observations were equally split over two observatory roll angles spanning $\sim$10 degrees. Each roll includes 2 detectors (NRS1 and NRS2), 33 groups, 1 integration, and 16 dithers with NIRSpec's small cycling pattern totaling $\sim132$ min per roll. The 16 small dithers are chosen to improve the spatial sampling of the NIRSpec instrument. To minimize saturation effects from the bright host star ($K\approx2.2$), the planet is placed near the center of the IFU, and the PSF core of the star is outside the FoV.

\subsubsection*{NIRSpec data reduction}
Starting with the raw uncalibrated detector images, we first performed reductions using stage 1 and stage 2 of the JWST Science Calibration Pipeline v2.0.1 \cite{bushouse2026}. We used the Calibration Reference Data System (CRDS) software version 13.1.16 ($\texttt{CRDS\_VER}$), and the CRDS context jwst{\_}1535.pmap ($\texttt{CRDS\_CTX}$). From stage 2, the resulting flux-calibrated detector images (\texttt{*\_cal.fits}) are further processed using the open-source \texttt{breads} package \cite{Ruffio2024} to subtract the starlight and obtain the planet detection map (Fig.~\ref{fig:detection}) and spectrum (Fig.~\ref{fig:spectrum}). A detailed description of the \texttt{breads} reduction process for JWST can be found in previous work \cite{Ruffio2024, Ruffio2026}, and we briefly summarize the main steps and highlight a few updates.

In raw NIRSpec frames, the planet is a few times fainter than the stellar PSF diffraction features. To detect the planet, we fit a joint forward model of the planet and diffracted starlight directly in the detector images. For the star, we first derive a continuum-normalized stellar spectrum from the science data after masking the planet location. This continuum-normalized stellar spectrum is later combined with a flexible 1D spline model following detector rows to model the starlight in the data. The planet is modeled using a model PSF combined with a Sonora Flame Skimmer model \cite{2026arXiv260806454M} with $\Teff=275~$K, $\logg=4.0$, $\log{K_{\rm zz}}=4.0$, [M/H]=0.5 dex, and C/O=0.458. 

In previous studies using the \texttt{breads} framework, the model PSF was computed from a numerical model of JWST/NIRSpec with STPSF \cite{Perrin2014SPIE.9143E..3XP}. However, given the  high S/N of $\epsilon$ Ind Ab (detected at $\approx600~\sigma$, Fig.~\ref{fig:detection}), inaccuracies in the STPSF model lead to systematics in the extracted flux of the planet. In this work, we use an empirical PSF model derived from observations acquired by the JWST PSF calibration program GO 3399 (PI: M. Perrin) as follows. The program GO 3399 observed two stars using a sequence of 16 small cycling dithers. One of the stars is faint enough to not saturate (2MASS J17571324+6703409; K = 11.15). It is used to probe the core of the PSF. The second star is brighter so its central core saturates (HD 163466; K = 6.33); however, it is used to obtain higher S/N data for the fainter tails of the PSF. The 16 dithers of each star were combined to create a super-sampled effective PSF (ePSF) \cite{Anderson2000PASP..112.1360A} by fitting a 3D (x, y, $\lambda$) spline model to the point cloud. The sampling of the ePSF is 20 mas compared to the original 100 mas of the NIRSpec spatial pixels. In this context, the point cloud refers to the ensemble of illuminated pixels on the NIRSpec detectors, which irregularly samples the NIRSpec PSF. The best-fit 3D spline model of both stars were then combined to create the final high-S/N empirical model of the NIRSpec PSF in the F290LP/G395H observing mode. The resulting model PSF provides a more accurate spectral extraction compared to STPSF. 

By fitting the forward model of the star and planet across the IFU in each spatial location, we estimate the planet's detection S/N at each pixel. We normalize the S/N maps to achieve a standard deviation of unity in the absence of planet signal. Combining all wavelengths, we obtain a planet detection S/N of $\approx600$, shown in Fig.~\ref{fig:detection}.

To extract the planet spectrum, we apply the spline-based high-pass filter to fit and subtract the starlight from everywhere in the detector. Note that this step effectively removes the continuum of the planet while preserving its high-resolution spectral lines. 
We then fit the empirical PSF model at each wavelength over a grid of right ascension (RA) and declination (Dec) to construct an initial planet flux cube. Each element of the cube stores the fitted flux for a planet placed at that position and wavelength. This initial cube uses sky coordinates derived from the World Coordinate System (WCS), without the correction for the wavelength-dependent centroid \cite{Ruffio2024}. We use this initial cube to locate the planet and measure this wavelength-dependent drift before repeating the cube extraction with corrected coordinates.

Unlike previous NIRSpec datasets, where the IFU is centered on the star \cite{Ruffio2026, Xuan2026b}, in these observations the star is outside the FoV. In this case, we found that determining the wavelength-dependent centroid shift of the PSF \cite{Ruffio2024} using the stellar diffraction features is challenging. As the planet is detected with a high S/N, we instead solve for the chromatic centroid shift using the planet signal itself. In short, we cross-correlate the same Sonora Flame Skimmer model mentioned above with the high-pass-filtered planet spectrum at each spatial location in the spectral cube, in contiguous wavelength intervals of $0.1~\mu$m. We determine the planet position from the peak of the cross-correlation function in each interval and fit first-order polynomials to the measured $\Delta\mathrm{RA}$ and $\Delta\mathrm{Dec}$ offsets as a function of wavelength. This procedure recovers a $\approx15$ mas centroid shift, predominantly in the $\mathrm{Dec}$ direction. The size of the centroid shift is found to be consistent between the two NIRSpec rolls. We apply these wavelength-dependent polynomial corrections to the WCS-derived coordinates to build the corrected planet flux cube. We then cross-correlate the corrected cube with the Sonora Flame Skimmer template to determine the final planet position. The final 1D planet spectrum is extracted at the peak of the resulting cross-correlation map. We estimate the flux uncertainties and covariance matrix of the planet spectrum from starlight residuals within an annulus centered on the host star at the same projected separation as the planet, after masking out a $0.3^{\prime\prime}$ radius region around the planet's position. 

\subsubsection*{Atmospheric retrieval setup}
Here we describe the process of inferring atmospheric parameters from the observed planet spectra. We use the radiative transfer code \texttt{petitRADTRANS} \cite{molliere_petitRADTRANS_2019, Nasedkin2024_PRT} to calculate synthetic planet templates, which are compared to the observed spectra in a Bayesian retrieval analysis using nested sampling. The atmospheric retrieval framework used in this work closely follows previous work \cite{Xuan2026b}, which we refer to for details. 

For the NIRSpec spectra, we use the line-by-line opacity sampling mode with $R=100,000$ to construct a model from $2.85-5.30~\mu$m. For photometry, we use the correlated-k sampling mode with $R=300$ to generate a model covering the remaining wavelengths. A separate model at $R=300$ is used for the photometry to reduce the computational cost. The parameters we fit for include the planet mass and radius, radial velocity, temperature profile, chemical abundances, and cloud structure, as well as nuisance parameters for the spectral resolution and error inflation terms. All fitted parameters and their priors are summarized in Extended Data Table~\ref{tab:param_prior}.

We sample the posterior distributions with the nested sampling package \texttt{pymultinest} \cite{Buchner2014, Feroz2019}. We use 1000 live points and stop sampling when the estimated contribution of the remaining prior volume to the total evidence is $<1\%$. The Bayesian evidence from each fit allows us to calculate the Bayes factor $B$, which quantifies the relative probability of different models.

For each synthetic spectrum from \texttt{petitRADTRANS}, we apply the following steps before comparing the model with the observed NIRSpec spectra. First, we apply a radial velocity shift to the model to align the model spectra with the data. Second, we apply instrumental broadening to the model and resample the model to the data wavelengths. We fit the NIRSpec resolving power as a linear function of wavelength \cite{Xuan2026b}, parameterized here by its values at the blue and red endpoints, $r_{\rm blue}$ and $r_{\rm red}$. Third, we apply the same spline-based high-pass filtering on the model that was applied on the data. This means we remove the best-fit spline from the model using the same node positions in the wavelength direction. After these steps, we compute the log likelihood as follows:

\begin{equation}
    \ln \mathcal{L} = -\frac{1}{2} \left( R^\intercal C^{-1} R + n \ln(2\pi) + \ln\det(C) \right)
\end{equation}

where $R$ is the residual array, $C$ is the covariance matrix, $n$ is the number of data points, and ${\rm det}(C)$ is the determinant of the covariance matrix. Here, $C$ has been multiplied by a factor of $e_{\rm mult}$, which is the error inflation term that we fit. Due to the significantly different flux levels of the planet in NRS1 and NRS2 wavelengths, we fit two separate error inflation terms, one for NRS1 and one for NRS2.

To constrain the continuum flux of the planet, we jointly model the literature photometry with the NIRSpec spectra (Extended Data Fig.~\ref{fig:nirspec_and_phot}). The photometry data span 4 to 25 $\mu$m, and are obtained by JWST/NIRCam and JWST/MIRI \cite{Matthews2024, Matthews2026, Sanghi2026}. In the Bayesian atmospheric retrieval, the log likelihoods from the photometry and NIRSpec components are added.

To parametrize the thermal structure, or pressure-temperature (P-T) profile, of the planet's atmosphere, we use the gradient-based profile \cite{Zhang2023_AFLep}, which fits for $(d\ln{T}/d\ln{P})$ values at fixed pressure nodes and a reference temperature, $T_{\rm ref}$, which we set at 0.1 bars. We define 13 pressure nodes spanning $10^{-6}$--$10^{2.7}$ bar to resolve the pressure range relevant to the planet's emission contribution function. The exact node locations are $\log_{10}(P/{\rm bar}) = [-6,\,-5,\,-4,\,-3,\,-2,\,-1,\,0,\,\allowbreak 0.7,\,1.0,\,1.3,\,1.7,\,2.0,\,2.7]$. The node spacing is finer near the photosphere and in the deep atmosphere. The denser grid in the deep atmosphere prevents interpolation across wide pressure ranges from imposing an artificially rigid P–T profile.

We consider three different cloud models: clear, gray clouds, and the EddySed model \cite{ackerman_Precipitating_2001} with water clouds implemented in \texttt{petitRADTRANS} \cite{molliere_Retrieving_2020}. Given the cold temperature of $\epsilon$ Ind Ab, water clouds are expected to be the most relevant cloud species at near-infrared wavelengths \cite{Lacy2023, Matthews2024}. Previous work modeling $\epsilon$ Ind Ab's photometry data has also hinted at the presence of water clouds in its atmosphere \cite{Matthews2026, Sanghi2026}.

We consider two different chemical setups for the retrievals. The first setup is known as ``free retrievals'', where the mass-mixing ratio of each species is fitted as a separate parameter and assumed to be constant with pressure. The advantage of this approach is mainly flexibility, as it assumes less prior knowledge about the chemistry. We also tested an alternative free retrieval where we allow the gas abundance of H$_2$O (and H$_2^{18}$O, H$_2^{17}$O) to decrease as a power law towards the top of the atmosphere \cite{Kuhnle2025}. Such water depletion is motivated by the expected formation of water clouds, and this alternative free retrieval is used to test whether the NIRSpec data find evidence of water-depletion. 

In the second setup, which is known as chemically-consistent retrievals, the individual mass-mixing ratios are determined by elemental abundance parameters and an equilibrium chemistry network \cite{Xuan2026, deRegt2026}. Compared to free retrievals, this approach is more physically-motivated. In this work, we use the equilibrium chemistry code \texttt{FastChem} \cite{Kitzmann2024}, specifically version 4.0.3 (\url{https://newstrangeworlds.github.io/FastChem/}). For each iteration of the retrieval, we run a new \texttt{FastChem} calculation given the P-T profile, and four fitted parameters: [C/H], [O/H], [N/H], and [S/H]. We include rainout condensation \cite{Kitzmann2024}. For computational efficiency, we use a simplified \texttt{FastChem} network \cite{deRegt2026} with 46 gas-phase species and 25 condensates (nearly an order of magnitude less than the full \texttt{FastChem} network). We verified that our reduced network reproduces the major gas-phase abundances relevant for $\epsilon$ Ind Ab-like P-T profiles, and runs $20\times$ faster. We adopt the solar elemental abundances \cite{Asplund2009} for the reported [C/H], [O/H], [N/H], and [S/H]. 

To account for disequilibrium chemistry in the second setup, we fit the vertical eddy diffusion coefficient, $K_{\rm zz}$, and apply the quench approximation \cite{Zahnle_methane_2014}. We calculate the quench pressures by equating the relevant chemical timescale to the vertical mixing timescale
\begin{equation}\label{eq:mix}
    \tau_{\rm mix}
    =
    \frac{L^2}{K_{\rm zz}}.
\end{equation}
Here, $L$ is characteristic mixing length and is typically assumed to be the pressure scale height $H = \frac{k_B T}{\mu\, m\, g}$ ($k_B$: Boltzmann constant; $\mu$: mean molecular weight in amu; $m$: atomic mass unit; $g$: surface gravity). We discuss the choice of $L=H$ further in the `Vertical eddy diffusion coefficient' section. There are three relevant quench pressures for $\epsilon$ Ind Ab: one for CO, CH$_4$, and H$_2$O, one for NH$_3$ and N$_2$, and one for CO$_2$. The abundances of CO, CH$_4$, H$_2$O, CO$_2$, and NH$_3$ at $P<P_{\rm quench}$ are set equal to their abundances at the respective quench points. The chemical reaction timescales are taken from previous work \cite{Zahnle_methane_2014, Mukherjee2022}. For CO$_2$, we use the disequilibrium H$_2$O and CO profiles to calculate a pseudo-equilibrium CO$_2$ profile, to which we then apply the CO$_2$ quench approximation \cite{Wogan2025}.

Since $\epsilon$ Ind Ab is cold enough for volatile condensation to occur, we additionally limit the post-quench gas abundances such that the partial pressure of each condensable species does not exceed its saturation vapor pressure. We apply saturation caps for all volatile species (H$_2$O, CH$_4$, CO, NH$_3$, H$_2$S, and CO$_2$) using vapor pressure relations \cite{Fray2009}. We note that the saturation caps are applied only in the chemically-consistent retrieval.

We detect several minor isotopologues in the NIRSpec spectrum of $\epsilon$ Ind Ab. In the retrievals, we fit the mass fraction of the major isotopologue relative to the minor isotopologue as a log mass fraction ratio. For example, the parameter for CH$_3$D is
\begin{equation}
    \log_{10}\left(\frac{X_{\rm CH_4}}{X_{\rm CH_3D}}\right),
\end{equation}
where $X_i$ is the mass fraction of species $i$. The corresponding number ratio is
\begin{equation}
    \frac{n_{\rm CH_4}}{n_{\rm CH_3D}}
    =
    \frac{X_{\rm CH_4}}{X_{\rm CH_3D}}
    \frac{\mu_{\rm CH_3D}}{\mu_{\rm CH_4}},
\end{equation}
where $\mu_i$ is the molecular mass. For the other isotopologues besides CH$_3$D, the number ratios from the equation above yields the isotope ratios of $^{12}$C/$^{13}$C from CO and CH$_4$, $^{16}$O/$^{18}$O from CO and H$_2$O, and $^{16}$O/$^{17}$O from H$_2$O. To obtain the D/H ratio from CH$_4$, the molecular number ratio needs to be converted to an atomic ratio as CH$_4$ contains four H atoms and CH$_3$D contains one D atom and three H atoms:
\begin{equation}
    ({\rm D/H})_{\rm CH_4}
    =
    \frac{n_{\rm CH_3D}}
    {4n_{\rm CH_4} + 3n_{\rm CH_3D}}.
\end{equation}

The line opacities included in the retrievals are summarized in Extended Data Table~\ref{tab:line_opacities}. For continuum opacities, we include the collision-induced absorption (CIA) from H$_2$-H$_2$ and H$_2$-He. Given the large number of isotopologues and molecules detected in $\epsilon$ Ind Ab's spectrum, we calculate new opacity tables for all line species using \texttt{line-racer} \cite{Hagele2026}. In this process, we noted the default opacity tables in \texttt{petitRADTRANS} versions 3 or below use a 1 dex pressure spacing, which created noticeable interpolation errors. Therefore, for the new calculations, we compute opacities on a grid of $\log_{10}(P/\text{bar})$ from $-7$ to $3$ in steps of 1~dex, with finer steps of $1/3$~dex between $\log_{10}(P/\text{bar})=-2$ and $1$, near the peak of the emission contribution function.

\subsubsection*{Atmospheric retrieval results}

\begin{figure}[h]
\centering
\makebox[\linewidth][c]{%
  \includegraphics[width=0.9\linewidth]{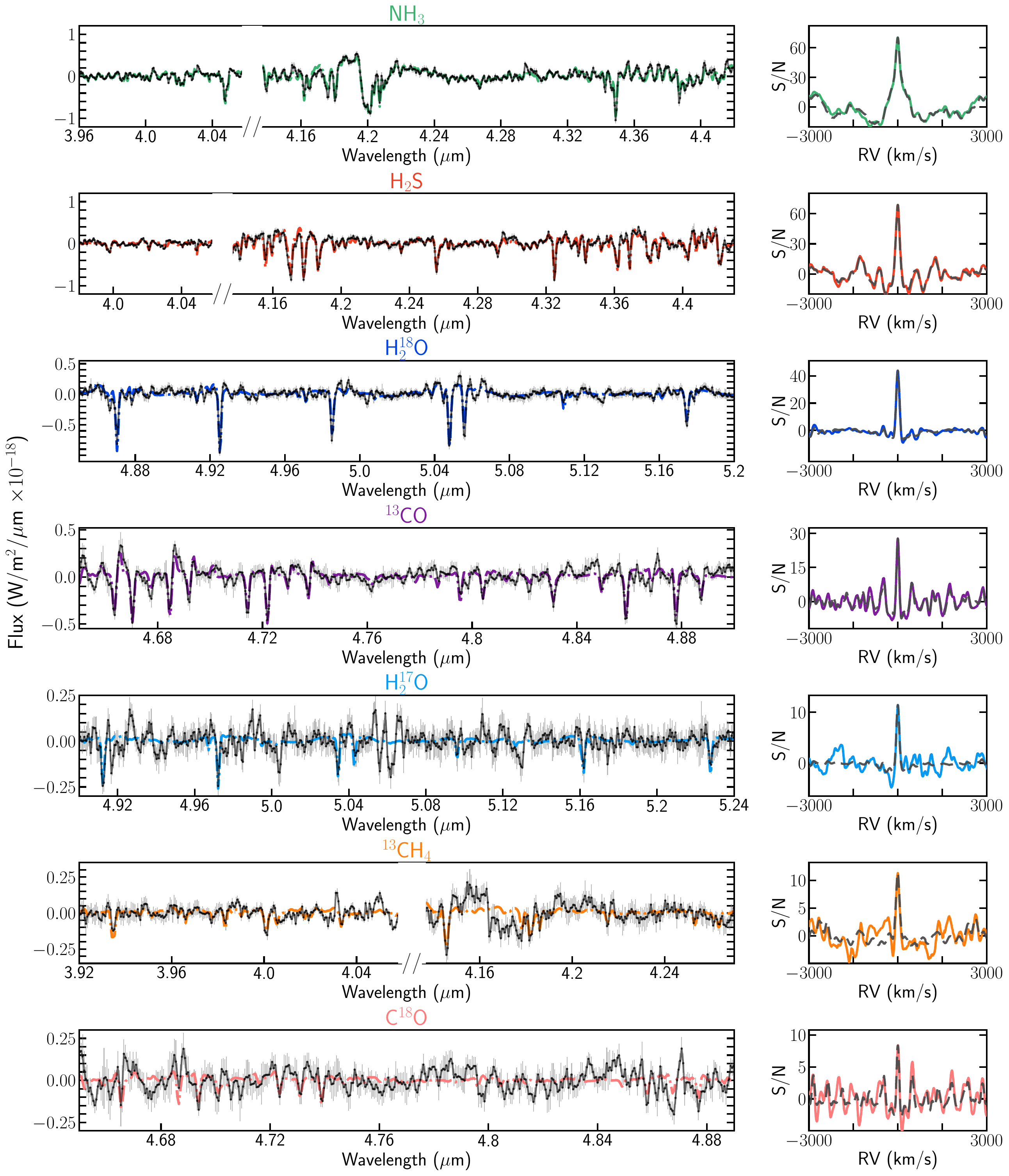}%
}
\caption{\textbf{Detection of NH$_3$,  H$_2$S, H$_2^{18}$O, $^{13}$CO, H$_2^{17}$O, $^{13}$CH$_4$, and C$^{18}$O in $\epsilon$ Ind Ab.} In each row, the left panel shows the data residuals after subtracting the corresponding leave-one-out retrieval (black). The colored dashed lines show the molecular template. On the right panels, the cross-correlation function between the data residuals and molecular templates are plotted in color. This provides an estimate of the detection S/N for each species. The dashed gray line shows the auto-correlation function of the molecular template. 
}
\label{fig:other_detections}
\end{figure}

In the NIRSpec spectra of $\epsilon$ Ind Ab, we detect 12 molecular isotopologues spanning six parent molecules: $^{12}$CH$_4$, $^{12}$CH$_3$D, $^{13}$CH$_4$, H$_2^{16}$O, H$_2^{18}$O, H$_2^{17}$O, $^{12}$CO, $^{13}$CO, C$^{18}$O, CO$_2$, H$_2$S, and NH$_3$. To validate these detections, we run leave-one-out retrievals where, each time, one isotopologue is left out from the line opacities. Next, we compare the residuals of the leave-one-out models with single-species templates \cite{Xuan2026}. For example, the residuals of the leave-one-out model without $^{12}$CH$_3$D would contain $^{12}$CH$_3$D lines if the data shows evidence of this isotopologue (as shown in panel b of Fig.~\ref{fig:spectrum}). From cross-correlating the data residuals of the model without $^{12}$CH$_3$D, and a $^{12}$CH$_3$D-only template, we calculate a cross-correlation function (CCF) S/N \cite{Ruffio2017}. From the CCF analysis, CH$_3$D is detected at $48\sigma$, and the five carbon and oxygen isotopologues are detected at significances ranging from $8\sigma$ to $44\sigma$, as shown in Extended Data Fig.~\ref{fig:other_detections}. The CCF S/N values for NH$_3$ and H$_2$S are $\approx 70$. We do not perform this CCF exercise for species such as $^{12}$CO and H$_2^{16}$O whose lines are clearly evident in the data. 

\begin{figure}[h]
\centering
\makebox[\linewidth][c]{%
  \includegraphics[width=1\linewidth]{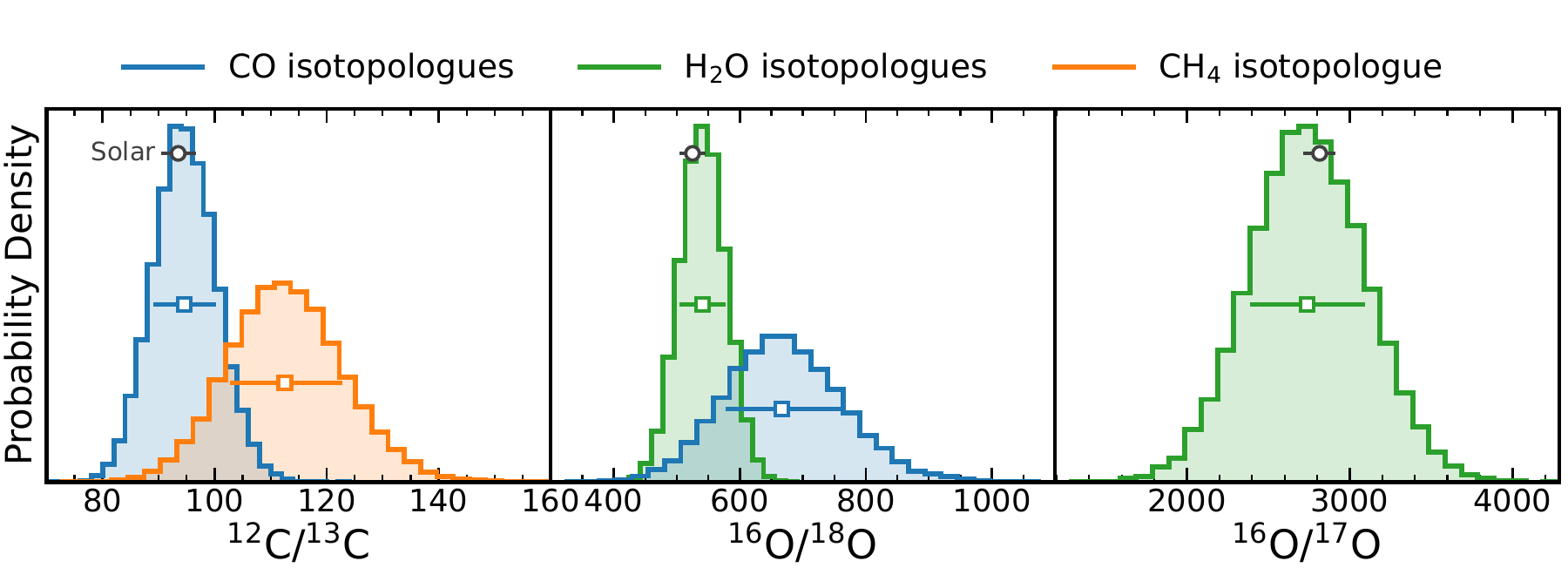}%
}
\caption{\textbf{Carbon and oxygen isotope ratios measured from different isotopologues in $\epsilon$ Ind Ab.} CO is blue, H$_2$O is green, and CH$_4$ is orange. In each panel, the circle marker denotes the solar ratios \cite{Lyons2018}. $^{12}$C/$^{13}$C is measured from both $^{13}$CO and $^{13}$CH$_4$, whereas $^{16}$O/$^{18}$O is measured from both C$^{18}$O and H$_2^{18}$O. The isotopic ratios are consistent with the solar value at the $<1\sigma$ level.
}
\label{fig:c_o_iso}
\end{figure}

The isotopic ratios of $^{12}$C/$^{13}$C, $^{16}$O/$^{18}$O, $^{16}$O/$^{17}$O from CO, CH$_4$, and H$_2$O are shown in Extended Data Fig.~\ref{fig:c_o_iso}. Unlike the case of D/H, the planet's carbon and isotopic isotopic ratios measured from different molecules are consistent with the corresponding solar ratios \cite{Lyons2018} at the $<1\sigma$ level. In addition, the isotopic ratios measured from different molecules (i.e. $^{13}$CH$_4$ and $^{13}$CO; H$_2^{18}$O and C$^{18}$O) are also consistent with each other. 

We also searched for additional species in the NIRSpec spectrum. We tested but found no significant evidence for HDO, C$^{17}$O, $^{13}$CO$_2$, PH$_3$, C$_2$H$_2$, HCN, SiH$_4$, $^{13}$C$^{18}$O, GeH$_4$, and AsH$_3$. We note that the non-detection of HDO implies a $3\sigma$ upper limit of $\rm (D/H)_{H_2O} < 6\times10^{-5}$, which is consistent with the measured $\rm (D/H)_{CH_4}$ value, 

To place the planet's heavy element enrichment into context, we need to know the host star's abundances. Two recent studies report C, O, and S abundances for the star $\epsilon$ Ind A \cite{Kolecki2022, Baburaj2026b}, finding slightly sub-solar abundances. The values from these studies agree at the $1\sigma$ level, so we adopt the newer values \cite{Baburaj2026b}, which are ${[\rm C/H]_\star}=-0.22\pm0.07$, ${[\rm O/H]_\star}=-0.14\pm0.08$, and ${[\rm S/H]_\star}=-0.15\pm0.18$. There is no literature measurement for [N/H] in $\epsilon$ Ind A, so we assume ${[\rm N/H}]_\star = [{\rm M/H}]_\star = -0.12\pm0.08$ \cite{Baburaj2026b}. 

In the chemically-consistent retrievals, we directly retrieve [C/H], [O/H], [N/H], and [S/H], and find that $\epsilon$ Ind Ab is enriched by $\approx5\times$ stellar in C, O, N, and S (Fig.~\ref{fig:elements}). From the free retrievals, we can add up the retrieved gas-phase molecular abundances to estimate [C/H], [O/H], and [S/H]. This yields ${\rm C/H}=4.9\pm1.0$, ${\rm O/H}=4.1^{+0.9}_{-0.7}$, and ${\rm S/H}=4.5^{+2.3}_{-1.5}$ $\times$ stellar, which are slightly lower by $\sim20\%$ but consistent within $1\sigma$ of the results from chemically-consistent retrievals. We do not quote N/H from free retrievals as they do not account for N$_2$, which constitutes a substantial part of the nitrogen budget even at $\Teff\approx275~$K. In addition, the free retrieval neglects elements sequestered in condensates. Therefore, the resulting elemental abundances from the two retrieval frameworks are not directly comparable. Nevertheless, the broadly consistent [C/H], [O/H], and [S/H] values suggest that gas-phase CH$_4$, H$_2$O, CO, CO$_2$, and H$_2$S are the dominant carriers of C, O, and S in $\epsilon$~Ind~Ab's atmosphere, as expected from self-consistent atmospheric models \cite{2026arXiv260806454M}.

We measure a planetary radial velocity of $-45.6\pm0.9~$\kms at Oct 25 (UT) 2025. This value is close to the host star RV of $-40.43\pm0.13~$\kms reported by the Gaia DR3 data release \cite{Gaia2023_DR3}, which has a reference epoch of 2016.0. Since the change in the host star's RV over two decades due to a combination of orbital motion and perspective acceleration is $<0.1~$\kms \cite{Sanghi2026}, our measurement would imply a relative RV (defined as planet RV - stellar RV) of $-5.2\pm0.9~$\kms, implying the planet is currently moving towards Earth in its orbit. This is consistent with the prediction from the latest orbit fits of the planet \cite{Sanghi2026}.

\begin{figure}[h]
\centering
\includegraphics[width=0.49\linewidth]{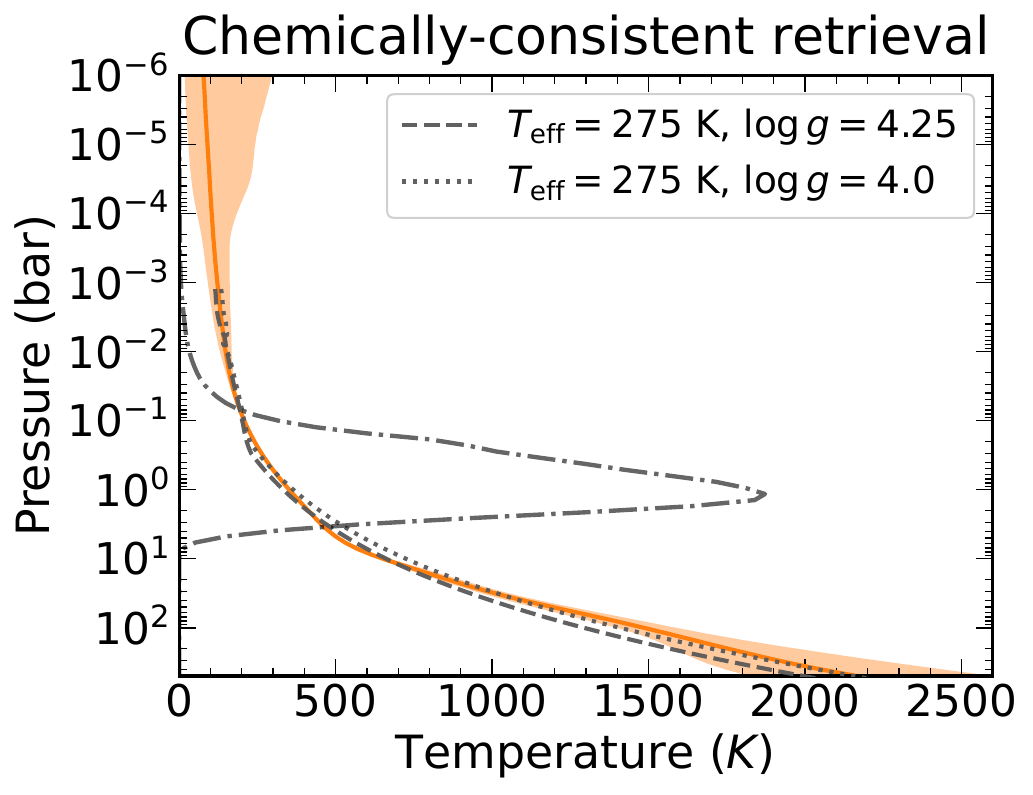}
\hfill
\includegraphics[width=0.49\linewidth]{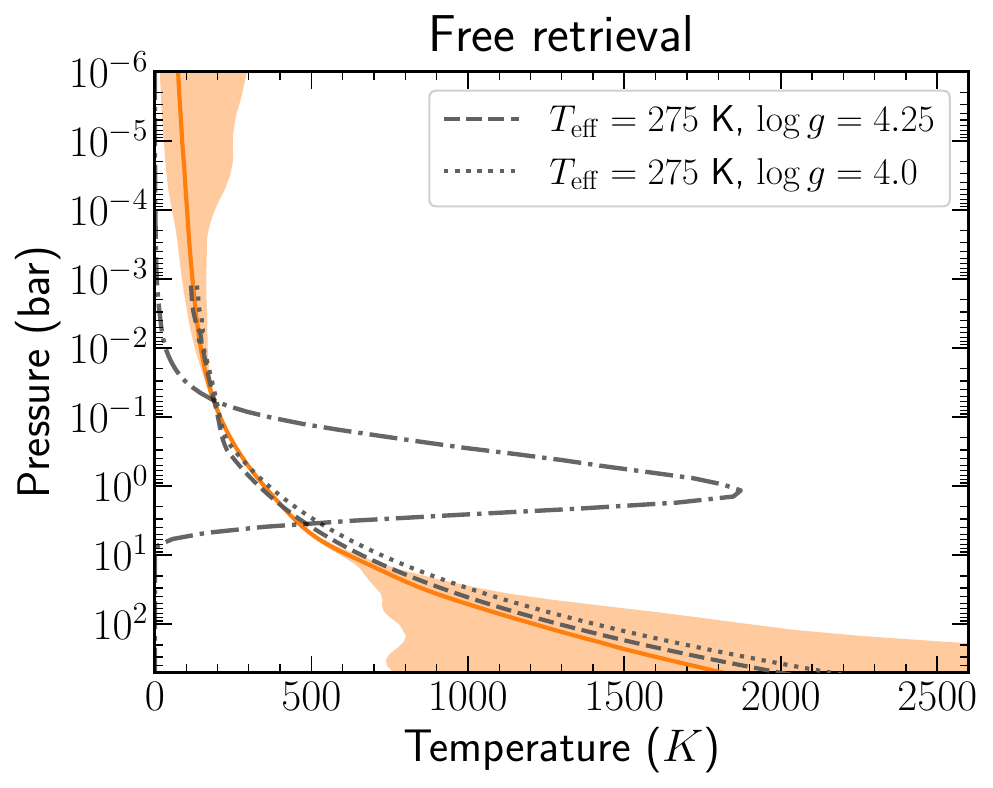}
\caption{\textbf{Retrieved P-T profiles for $\epsilon$ Ind Ab.} Profiles from the chemically-consistent retrieval (left) and free retrieval (right). The orange line is the median retrieved P-T while the orange shade denotes the $2\sigma$ interval on the retrieved profiles. In both panels, we overplot P-T profiles from the Sonora Flame Skimmer model \cite{2026arXiv260806454M} with $\Teff=275~$K, C/O=0.45, [M/H]=0.5 dex, $\log{K_{\rm zz}}=7.0$, and two values of surface gravity: log($g$)=4.25 (dashed line) and log($g$)=4.0 (dotted line). We show the weighted emission contribution function probed by NIRSpec using the dashdot line on the pressure axis. From $P\approx10-100$ bars, the profiles are more constrained in the chemically-consistent retrieval since the deeper atmosphere and photosphere are connected via quenching of the chemical equilibrium abundances (see `$P$--$T$ profile and bulk properties').}
\label{fig:pt}
\end{figure}

The retrieved $P$--$T$ profiles from the chemically-consistent and free retrievals are shown in Extended Data Fig.~\ref{fig:pt}, and broadly match self-consistent $P$--$T$ profiles from Sonora Flame Skimmer \cite{2026arXiv260806454M}. The NIRSpec spectra are mainly sensitive to regions between $10^{1}$ to $10^{-2}$ bars in the atmosphere, as indicated by the weighted emission contribution function. At pressures of $\approx10-100$ bar, near the inferred CO$_2$, CO/CH$_4$/H$_2$O, and NH$_3$ quench points, the chemically-consistent retrieval yields a narrower range of P-T profiles compared to the free retrieval. This is because in the chemically-consistent retrieval, the photosphere and deeper atmosphere are linked through quenching and the chemical equilibrium abundances. In other words, when quenching happens below the photosphere, the molecular abundances in the photospheric region are controlled by both the equilibrium abundances below and the $K_{\rm zz}$. 

To estimate the planet's bolometric luminosity, we integrate the model spectra from $0.3-50~\mu$m. To account for the remaining flux at longer wavelengths, we additionally integrate a blackbody spectrum normalized to match the model flux at $50~\mu$m and integrate it to an arbitrarily large wavelength of $10^6~\mu$m. We find the planet's bolometric luminosity to be $\lbollsun=-7.23\pm0.01$. Given the retrieved radius of $1.02\pm0.01~\Rj$, this implies $\Teff=277\pm2~$K. These values are very consistent with those derived previously by modeling only $\epsilon$ Ind Ab's $4-25~\mu$m photometry data with self-consistent atmospheric models \cite{Sanghi2026}.

By comparing the three clouds models -- clear, gray cloud, and EddySed with water clouds -- we find that the gray cloud model is strongly preferred over the other two with $\ln{B}\approx20$. The clear model PT profile shows an isothermal region around 10 bars and is highly discrepant with the self-consistent PT shapes, suggesting the clear model adjusts the PT profile to compensate for the lack of an gray opacity source from clouds. The clear and EddySed models also find a planet radius that is too small ($\approx0.9~\Rj$). While water clouds could suppress and alter the $4-5~\mu$m flux of cold planets like $\epsilon$ Ind Ab \cite{Morley2014, Lacy2023, Matthews2026}, the effect of clouds is generally much easier to discern in the spectral continuum. Additional work on improving PSF subtraction techniques for NIRSpec \cite{Ruffio2024} could allow the recovery of the planet's $3-5~\mu$m continuum spectra and further test the presence of water clouds.

As noted in the `Chemistry' section, we accounted for volatile condensation by applying a saturation cap in the chemically-consistent retrievals. However, in the free retrieval where we allow the gas-phase water abundance to decrease as a power law, we do not find evidence that such water depletion is preferred by the data. This is not surprising since for WISE 0855, only the MIRI/MRS spectra $>15~\mu$m were sensitive to the effects of water depletion \cite{Kuhnle2025}. A future study analyzing the JWST MIRI/MRS spectroscopy of $\epsilon$ Ind Ab from GO 8714 will explore the existence of water clouds in its atmosphere in more detail. 

We infer $\log_{10}(K_{zz}/{\rm cm^2\,s^{-1}})=5.6\pm0.4$ by combining the chemically-consistent retrievals of the two NIRSpec rolls. Note that this $\log_{10}(K_{zz})$ value assumes $L=H$, which is the commonly adopted convention for brown dwarf quenching calculations \cite{Visscher2011, Mukherjee2022}. If we instead adopt $L=0.14H$, as inferred for Jupiter \cite{smith_Estimation_1998}, the same mixing timescale would correspond to $\log_{10}(K_{zz}/{\rm cm^2\,s^{-1}})=3.9\pm0.4$. The corresponding quench pressures are $53\pm11$, $27\pm4$, and $90\pm22$ bars for CO--CH$_4$, CO$_2$, and N$_2$--NH$_3$, respectively. 

\subsubsection*{Methane--hydrogen fractionation correction for D/H}
From the retrievals, we measure the D/H in methane for $\epsilon$ Ind Ab. Because the atmosphere is dominated by H$_2$, the D/H in molecular hydrogen more closely traces the bulk atmospheric D/H. We therefore correct for chemical fractionation between CH$_4$ and H$_2$ using two complementary methods, described below.

We first perform an analytic calculation following an established method for Solar System giant planets \cite{Lecluse1996}. The methane--hydrogen enrichment factor is defined as
\begin{equation}
    f_{\rm CH_4-H_2}
    =
    \frac{({\rm D/H})_{\rm CH_4}}
         {({\rm D/H})_{\rm H_2}} .
\end{equation}
Deuterium is exchanged between methane and molecular hydrogen through
the net reaction
\begin{equation}\label{eq:iso_react}
    {\rm CH_4 + HD}
    \rightleftharpoons
    {\rm CH_3D + H_2}.
\end{equation}
In thermochemical equilibrium, the partitioning of D between methane and hydrogen can be described by the temperature-dependent fractionation factor $\alpha(T)$ \cite{Lecluse1996}. At temperatures exceeding $1200$ K, $\alpha(T)$ approaches unity so if the isotopic exchange reaction quenches in this regime $f_{\rm CH_4-H_2}$ would also be close to unity \cite{Lecluse1996}. As gas is mixed upward in the atmosphere, the isotopic exchange reaction becomes slower until its chemical timescale $\tau_{\rm chem}$ exceeds the vertical mixing timescale $\tau_{\rm mix}$. The CH$_3$D/CH$_4$ ratio is then approximately fixed at its equilibrium value at the quench point. To estimate $f$, we need to estimate the quench point of the isotopic exchange reaction \cite{Fegley1988}. Under the quench approximation, the enrichment in the observed atmosphere is
\begin{equation}
    f_{\rm CH_4-H_2}
    \simeq
    \alpha(T_{\rm quench}).
\end{equation}
To find $T_{\rm quench}$, we compare the isotope exchange timescale
\begin{equation}\label{eq:isochem}
    \tau_{\rm chem}
    =
    \left[k^{-}(T)n_{\rm H_2}\right]^{-1}
\end{equation}
with the vertical mixing timescale given by Equation~\ref{eq:mix}. In Equation~\ref{eq:isochem}, $k^{-}(T)$ is the isotopic exchange rate constant \cite{Lecluse1994, Lecluse1996} and $n_{\rm H_2}$ is number density of H$_2$, which we estimate using the H$_2$ volume-mixing ratio from \texttt{FastChem} calculations in the retrieval. 

From independent retrievals of the two NIRSpec rolls, we obtain $({\rm D/H})_{\rm CH_4}=(5.4\pm0.2)\times10^{-5}$. Using the inferred $K_{\rm zz}$ from the retrievals, we find that the isotopic exchange reaction quenches at $P=9\pm2$~bar and $T=620\pm50$~K, which gives $f_{\rm CH_4-H_2}\approx1.35$. For comparison, a similar calculation for Jupiter gives $f_{\rm CH_4-H_2}=1.25\pm0.12$ \cite{Lecluse1996}.

The analytic calculations above used the isotopic exchange rate constant, $k^{-}(T)$, inferred from experimental data with initial CD$_4$--H$_2$ mixtures \cite{Lecluse1994}. To verify the isotopic fractionation of methane within a full chemical system, we also ran a kinetics model incorporating the relevant C--H--O elementary reactions. We used the open-source photochemical kinetics model \texttt{VULCAN} \cite{Tsai2017,Tsai2021}, expanding its existing chemical network\footnote{\url{https://github.com/shami-EEG/VULCAN/blob/a68d7151b71f0ad32a7868c57f6ec7be3013a68d/thermo/SNCHO_photo_network_2025.txt}} to include the following deuterated species: D, HD, D$_2$, OD, D$_2$O, HDO, CH$_3$D, and CH$_2$D (see Extended Data Table~\ref{tab:reactions}). To ensure consistency, the \texttt{VULCAN} models were run using the median retrieved elemental abundances and median $P-T$ profile from the retrieval. For the relevant conditions in $\epsilon$ Ind Ab's atmosphere, the kinetics model shows that the dominant pathway for isotope exchange between molecular hydrogen and methane is 
\begin{align}
    {\rm CH_4 + H} &\rightarrow {\rm CH_3 + H_2} \nonumber\\
    {\rm CH_3 + HD} &\rightarrow {\rm CH_3D + H},
    \label{eq:path}
\end{align}
where the reaction between the methyl radical and HD controls the isotope exchange efficiency. 

\begin{figure}[h]
\centering
\includegraphics[width=\linewidth]{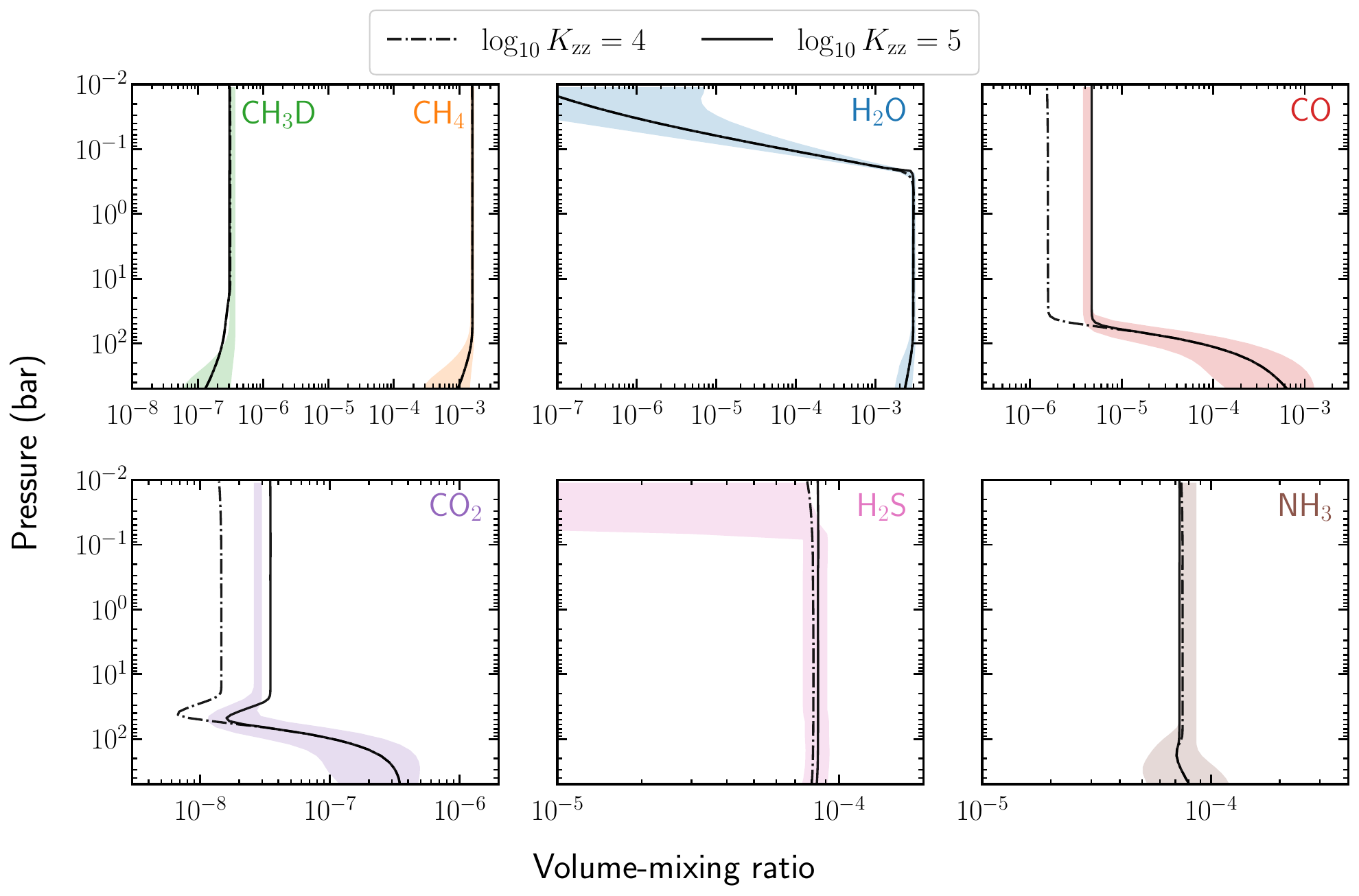}
\caption{\textbf{Volume-mixing ratio profiles for a subset of species detected in $\epsilon$ Ind Ab.} The shaded contours show the $2\sigma$ range from the chemically-consistent retrievals, whereas the black dashdot and solid lines show model abundance profiles from the \texttt{VULCAN} chemical kinetics code for two different $K_{\rm zz}$ values. Overall, the $\log_{10}(K_{\rm zz}/{\rm cm^2\,s^{-1}})=5$ \texttt{VULCAN} model matches the retrieved abundances well. Depending on the assumption of mixing length, the retrieval finds $\log_{10}(K_{\rm zz})=3.9-5.6$, which is consistent with the \texttt{VULCAN} value. The retrieval H$_2$O and H$_2$S abundance profiles decrease at low pressures due to condensation, which we implement using the saturation vapor pressure relations \cite{Fray2009}.}
\label{fig:vulcan}
\end{figure}

The kinetic results from \texttt{VULCAN} for two $K_{\rm zz}$ values are shown in Extended Data Fig.~\ref{fig:vulcan}, where we overplot the retrieved volume-mixing ratios for various species. The \texttt{VULCAN} model with $\log_{10}(K_{\rm zz}/{\rm cm^2\,s^{-1}})=5$ matches the retrieved abundances best. This $K_{\rm zz}$ value is consistent with the inferred $K_{\rm zz}$ from the retrievals. The equilibrium CH$_3$D/CH$_4$ ratio increases with decreasing temperature but varies only gradually with pressure near the inferred quench region. As a result, the fractionation factor $f_{\rm CH_4-H_2}$ is only weakly sensitive to $K_{\rm zz}$ here. For $\log_{10}(K_{zz}/{\rm cm^2\,s^{-1}})=5$, \texttt{VULCAN} places the isotope exchange quench point at approximately 18 bar and 809 K, giving $f_{\rm CH_4-H_2}=1.07$; for $\log_{10}(K_{zz}/{\rm cm^2\,s^{-1}})=4$, the corresponding values are 14 bar and 752 K, giving $f_{\rm CH_4-H_2}=1.10$.

In summary, the full kinetics calculation finds $f\approx1.10$, whereas the analytic calculation yields $f\approx1.35$. This demonstrates that for $\epsilon$ Ind Ab, isotopic fractionation only causes a modest $10-35\%$ difference between $(\mathrm{D/H})_{\rm CH_4}$ and $(\mathrm{D/H})_{\rm H_2}$. Neither value brings $(\mathrm{D/H})_{\rm H_2}$ close to the protosolar value. The difference between the $f$ values from the two calculations may arise from differences in the underlying chemical kinetics data. To conservatively account for model uncertainties, we choose to adopt a range of $f_{\rm CH_4-H_2}=1.10$--$1.35$. After applying the fractionation correction, we obtain $(\mathrm{D/H})_{\rm H_2} = (4.4\pm0.7)\times10^{-5}$, which represents the bulk atmospheric D/H of $\epsilon$ Ind Ab.

\subsubsection*{Modeling the atmospheric composition of $\epsilon$ Ind Ab}
We use the open-source composition framework \texttt{accomp} \cite{Chachan2023, Chachan2026} to relate the composition of $\epsilon$ Ind Ab to its building blocks. The code is available at \url{https://github.com/y-chachan/accomp}. In brief, this framework leverages measurements of volatile and refractory elements in a giant planet's atmosphere to deduce how the planet acquired its metals and the quantity of metals accreted via different sources such as solids, metals in local gas, or metal-enhanced disk gas or solids through processes such as radial drift of pebbles. The disk thermal structure is adopted from previous work \cite{Ida2016} with specific choices for stellar mass dependence of disk temperature \cite{Chachan2023b}. In the outer part of the disk where irradiation sets the disk temperature, our adopted temperature profile is $T_{\rm disk} \approx 43 \, (r/16 {\rm au})^{-3/7}$ K.

For $\epsilon$ Ind Ab, sulfur is a tracer of refractory solids as the planet is far out enough ($16\pm2$ AU) that all of it must have been in solid form at the planet's formation location \cite{Kama2019}. Thus, S/H informs the solid-to-gas accretion rate during envelope accretion and indicates that the planet acquired its metals through solid accretion. However, the planet's S abundance once normalized by the relatively imprecise stellar S abundance does not allow us to discriminate between different models. Instead, we rely on the more precisely measured C, O, and N abundances to put constraints on the planet's formation history. We vary the fraction of volatile species C, O and N that were present in disk solids at different locations to infer in which regions accretion would produce a composition that is compatible with the observed composition of $\epsilon$ Ind Ab. 

The near uniformity of C, O, N, and S enrichment suggests that the planet most likely accreted solids beyond the $N_2$ ($\sim 25 \, {\rm K}, \, 56$ au) or CO snowline ($\sim 30 \, {\rm K},~36$ au). Accretion of material between the CO$_2$ ($\sim 70 \, {\rm K},~5$ au) and CO snowlines would also match the observed composition. We proceed with the simplest model of accretion beyond the N$_2$ snowline, where all the metals are condensed into solids and their accretion produces uniform enrichment of C, O, N, and S. We note that the potential effects of additional processes, such as radial dust transport \cite{Booth2017,Schneider2021a,Ohno2026}, and local disk cooling due to self-shadowing \cite{OhnoUeda21}, are left for future studies. As discussed below, the simplest model of accretion beyond the N$_2$ snowline is also compatible with the accretion of cold and unprocessed D-rich material inherited from the parent molecular cloud. 

The planet's C, O, S, and N atmospheric abundances constrain how much metals it accreted during envelope accretion and provide a lower limit on its bulk metallicity and the metal mass in the planet \cite{Xuan2026, Chachan2026}. For super-Jupiters such as $\epsilon$ Ind Ab, most of the metals are expected to be in the envelope and the core's contribution is small in comparison \cite{Chachan2025b}. Assuming that the enrichment of Fe, Mg, and Si (i.e. other refractory species) follow that of S, we find that the planet's bulk metallicity is $0.043^{+0.007}_{-0.006}$. This translates to $89^{+18}_{-15}~M_\oplus$ of metals in the planet assuming a planet mass of $6.5 \pm 0.7~M_{\rm J}$. The planet's bulk metallicity is $\sim 4.5 \times$ the stellar $Z_\star = 0.0096$, which is commensurate with the super-stellar bulk metallicity found for the transiting giant planet population as well as several directly imaged super-Jupiters \cite{Chachan2025b, Xuan2026b}. 

The measured elemental abundances allow us to infer the D/H of accreted material needed to match the planet's D/H. We consider two sources of D-rich material: water ice and organics. In our composition model, a fraction $f_{\rm O, H_2O} = 0.45$ of the disk's O inventory is in the form of water and all of the planet's O comes from the accretion of solids. So, the amount $f_{\rm O, H_2O} \times x_{\rm O}$ comes from the accretion of water ice, where $x_{\rm O}$ is the measured enrichment of O. Each O atom brings 2 H atoms along with it, some of which are substituted with D. Similarly, a fraction $f_{\rm C, organics} = 0.5$ of C is assumed to be in organics in the disk and they contribute $f_{\rm C, organics} \times x_{\rm C}$ of the planet's observed enrichment assuming all of C comes from solid accretion. The amount of H each C atom brings along in organics is determined by (H/C)$_{\rm organics}$, and depends on how hydrogenated the organics are. The ratio (H/C)$_{\rm organics}$ can vary from $\sim 0.1 - 2$ for various organics based on observations of Solar System comets \cite{Fomenkova1994, Gail2017}. We adopt an intermediate (H/C)$_{\rm organics} = 1$ as our fiducial value because it represents the value of the (H/C) in complex organic matter, which constitutes most of the refractory carbon.

We calculate the planet's D/H from the D/H of its building blocks as
\begin{equation}
    {\rm (D/H)_p} = \frac{\sum_i n_{{\rm H+D}, i} {\rm (D/H)}_i / [1 + {\rm (D/H)}_i]}{\sum_i n_{{\rm H+D}, i} / [1 + {\rm (D/H)}_i]}, 
    \label{eq:DtoH_planet}
\end{equation}
where $n_{{\rm H+D}, i}$ and ${\rm (D/H)}_i$ are the total number of H and D atoms and the D/H, respectively, of species $i$. The $n_{{\rm H+D}, i}$ terms are normalized such that $\sum_i n_{{\rm H+D}, i} = 1$. The contributions of water and organics to H and D atoms are calculated as $2 f_{\rm O, H_2O} x_{\rm O} {\rm (O/H)_\star}$ and ${\rm (H/C)_{organics}} f_{\rm C, organics} x_{\rm C} {\rm (C/H)_\star}$, respectively. Stellar values for ${\rm (O/H)_\star} = 10^{8.55}/10^{12} = 10^{-3.45}$ and ${\rm (C/H)_\star} = 10^{8.21}/10^{12} = 10^{-3.79}$ \cite{Baburaj2026b} follow the same convention used for solar composition. 

Considering water ice and organics in isolation at first, we find that water ice needs to have a D/H $= 1.5^{+0.6}_{-0.5} \times 10^{-2}$ and organics need to have a D/H $= 5.4^{+2.1}_{-1.7} \times 10^{-2}$ for the planet's enrichment in C, O, and D to be mutually compatible. This can also be seen using an order-of-magnitude treatment (neglecting the $[1 + {\rm (D/H)}_i]$ terms in Equation~\ref{eq:DtoH_planet}), which yields (D/H)$_{\rm water} \sim (4.4 \times 10^{-5} - 2 \times 10^{-5}) / (2 \times 0.45 \times 5.2 \times 10^{-3.45}) \sim 1.5 \times 10^{-2}$ and (D/H)$_{\rm organics} \sim (4.4 \times 10^{-5} - 2 \times 10^{-5}) / (1 \times 0.5 \times 5.8 \times 10^{-3.79}) \sim 5 \times 10^{-2}$. For refractory organics, C/Si could be as high as 5.85, which corresponds to 70\% of C being in organics\cite{Bergin2015} (as opposed to our assumed 50\%), and H/C $\sim 1.5$ for pristine organics\cite{Paquette2021}. With these assumptions, our order-of-magnitude calculation yields a lower required value of (D/H)$_{\rm organics} \sim 2.5 \times 10^{-2}$. Extended Data Fig.~\ref{fig:DtoH_vs_OtoH_CtoH} shows the D/H obtained for the planet as a function of O/H and C/H for a range of D/H of water ice and refractory organics (H/C = 1), respectively. Accretion of material with Solar System-like D/H would require a much higher ($\sim200 \times$ stellar) C/H or O/H to match the observed D/H of $\epsilon$ Ind Ab. 

\begin{figure}[h]
    \centering
    \includegraphics[width=0.9\linewidth]{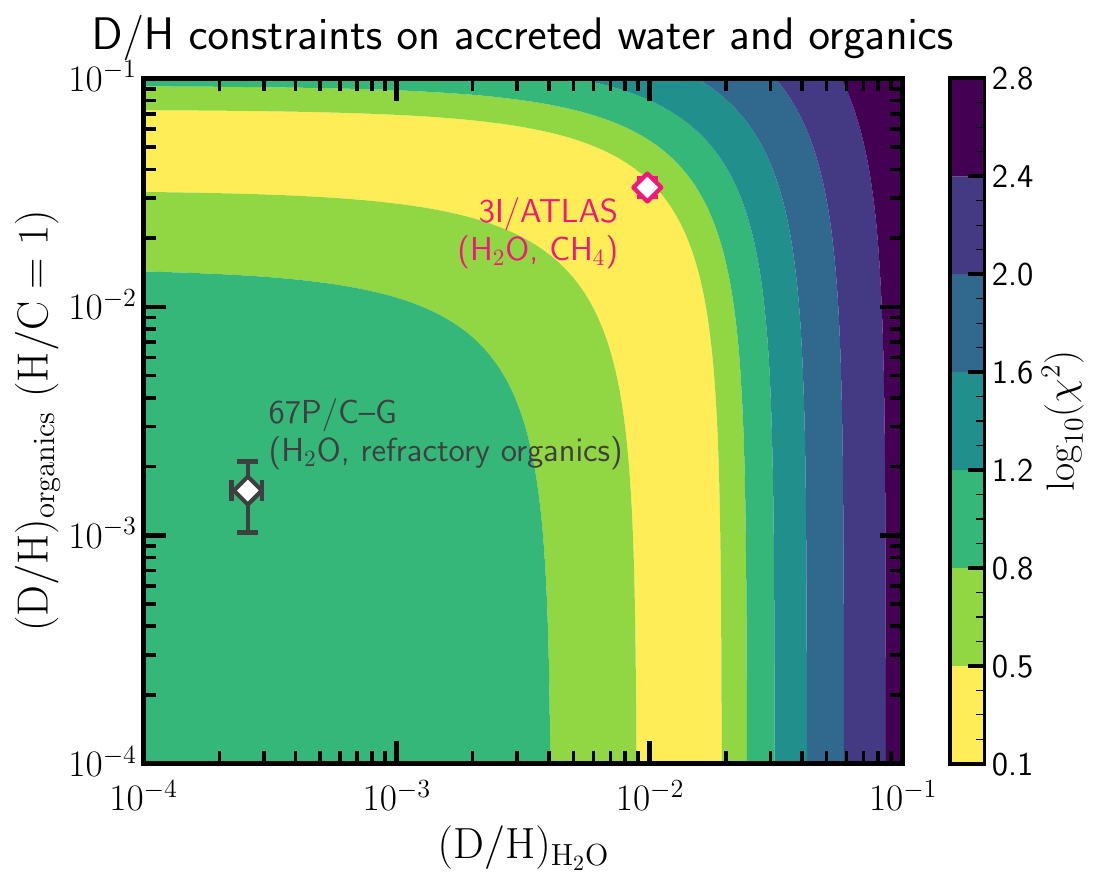}
    \caption{\textbf{The D/H required in water ice or refractory organics to match $\epsilon$ Ind Ab's composition is consistent with measurements of  the interstellar comet 3I/ATLAS.} The color map shows $\log_{10}(\chi^2)$ for combinations of water ice and refractory organic D/H that reproduce the planet's D/H, C/H, and O/H. Lower $\chi^2$ indicates better agreement. The black diamond shows the D/H measurements for 67P/C--G from water and refractory organics \cite{Mandt2024, Paquette2021}, while the pink diamond shows measurements for 3I/ATLAS from H$_2$O and CH$_4$ \cite{Cordiner2026, 2026arXiv260320445R}. If considered independently, water ice or refractory organics would need to have D/H of $\sim1.5\times10^{-2}$ or $\sim5\times10^{-2}$, respectively.}
    \label{fig:DtoH_water_organics}
\end{figure}

For a more comprehensive analysis, we create a grid of values for the D/H of water and organics (with H/C = 1) in the range $10^{-4} - 10^{-1}$. We calculate the planet's expected D/H over this grid and compare it to the measured D/H of the planet for the full set of retrieved abundance chains by calculating a $\chi^2$ value. Extended Data Fig.~\ref{fig:DtoH_water_organics} shows the $\chi^2$ map obtained for the D/H of water and organics to fit the planet's measured D/H. It agrees with the order-of-magnitude estimate above and suggests that, when considered independently, the D/H values of water and organics need to be $\sim 1.5 \times 10^{-2}$ and $\sim 5 \times 10^{-2}$, respectively, to match the planet's D/H given its O and C enrichment. 

\subsubsection*{Feasibility for accretion of deuterium-rich gas}
We also investigate whether the disk gas can become enriched in deuterium and heavy elements during disk dispersal, and whether this alternative scenario could explain the observed enrichment of D and metals in $\epsilon$ Ind Ab. The vertical gravitational force in the disk is so weak that the critical molecular/atomic mass of a species above which fractionation would occur is much larger than the mass of most relevant molecules (HD, CO, H$_2$O, N$_2$), i.e., the outflow itself is expected to be hydrodynamic and not lead to any fractionation\cite{Hunten1973, Guillot2006b}. Enrichment of disk gas in D and metals would only occur if these species are vertically settled relative to the background disk gas H$_2$ such that these species are depleted relative to H$_2$ at the wind-launching height above the midplane. This is equivalent to the concept of the homopause in planetary atmospheres above which the scale heights of different species depend on their molecular or atomic weight \cite{Hunten1973}. We can calculate the conditions under which such vertical separation would occur by comparing molecular diffusivity $D_m$ with turbulent diffusivity $D_t$. For heavier elements to settle relative to H$_2$, we need $D_m > D_t$. The molecular diffusivity of a binary mixture of H$_2$ and HD is given by $D_m = b_{\rm H_2, HD} / n$, where n is the total number density of the gas and $b_{\rm H_2, HD} \approx 5 \times 10^{17} \, T^{0.75}$ molecules cm$^{-1}$ s$^{-1}$ is the binary diffusion coefficient\cite{Mason1970, Genda2008} for HD and H$_2$ at temperature $T$. For turbulent diffusivity, we use $D_t = \alpha_t c_{\rm s} H$, where $\alpha_t$ is the Shakura-Sunyaev parameter, $c_{\rm s}$ is the sound speed, and $H$ is the disk scale height. The number density $n$ of the gas can be written as $n = \Sigma / (\sqrt{2 \pi} H \mu)$ for gas surface density $\Sigma$, and mean particle mass $\mu = 2.3 \times$ the mass of a proton, assuming a vertically isothermal and hydrostatic disk. For $\alpha_t = 10^{-4}$ and disk thermal structure adopted above ($T_{\rm disk} \approx 43 \, (r/16 {\rm au})^{-3/7}$ K), we find that vertical settling comes into play when
\begin{equation}
    \Sigma \leq \frac{b_{\rm H_2, HD} \sqrt{2 \pi} \mu}{\alpha_t c_{\rm s}} \approx 1.85 \times 10^{-5} \bigg( \frac{r}{16~{\rm au}} \bigg)^{-3/28} \bigg( \frac{\alpha_t}{10^{-4}} \bigg)^{-1} {\rm g ~cm^{-2}}.
\end{equation}
The required surface density is very weakly dependent on orbital distance and given that $\Sigma_{\rm disk}$ varies as $r^{-1}$ to $r^{-3/2}$, it is easier to meet this criterion further out in the disk. However, the pre-factor of $\sim 2 \times 10^{-5}$ implies a highly depleted nebula, e.g., a depletion of $\sim 5 \times 10^{-7}$ relative to the MMSN of Chiang \& Youdin\cite{Chiang2010} at 16 au. By the time a disk reaches such a low surface density, a giant planet is expected to have accreted almost the entirety of its mass. We verify this by calculating $\epsilon$ Ind Ab's mass doubling timescale using the maximum accretion rate from numerical simulations \cite{Choksi2023} and find that $t_{\rm double} = M_p/\dot{M} \approx 2$ Gyr, far longer than the disk lifetime. The planet would barely accrete any mass after the disk surface density has become low enough for vertical settling of deuterium and metals to matter. Therefore, accretion of deuterium-rich nebular gas is an implausible explanation for $\epsilon$ Ind Ab's elevated D/H. 

\subsubsection*{Benchmarking the retrieval framework with WISE 0855}
To benchmark our atmospheric retrieval framework, we ran a retrieval on the public NIRSpec/G395M spectrum of WISE 0855 from GO 2327 (PI A. Skemer). For this dataset, a value of $\rm (D/H)_{CH_4}=1.27^{+0.22}_{-0.19}\times10^{-5}$ was measured from CHIMERA retrievals \cite{Rowland2024, Line2015}. Using our free retrieval framework with \texttt{petitRADTRANS}, we fit the same WISE 0855 spectrum and obtain $\rm (D/H)_{CH_4}=1.27^{+0.15}_{-0.14}\times10^{-5}$, fully consistent with the literature value \cite{Rowland2024}.




\begin{figure}
    \centering
    \includegraphics[width=\linewidth]{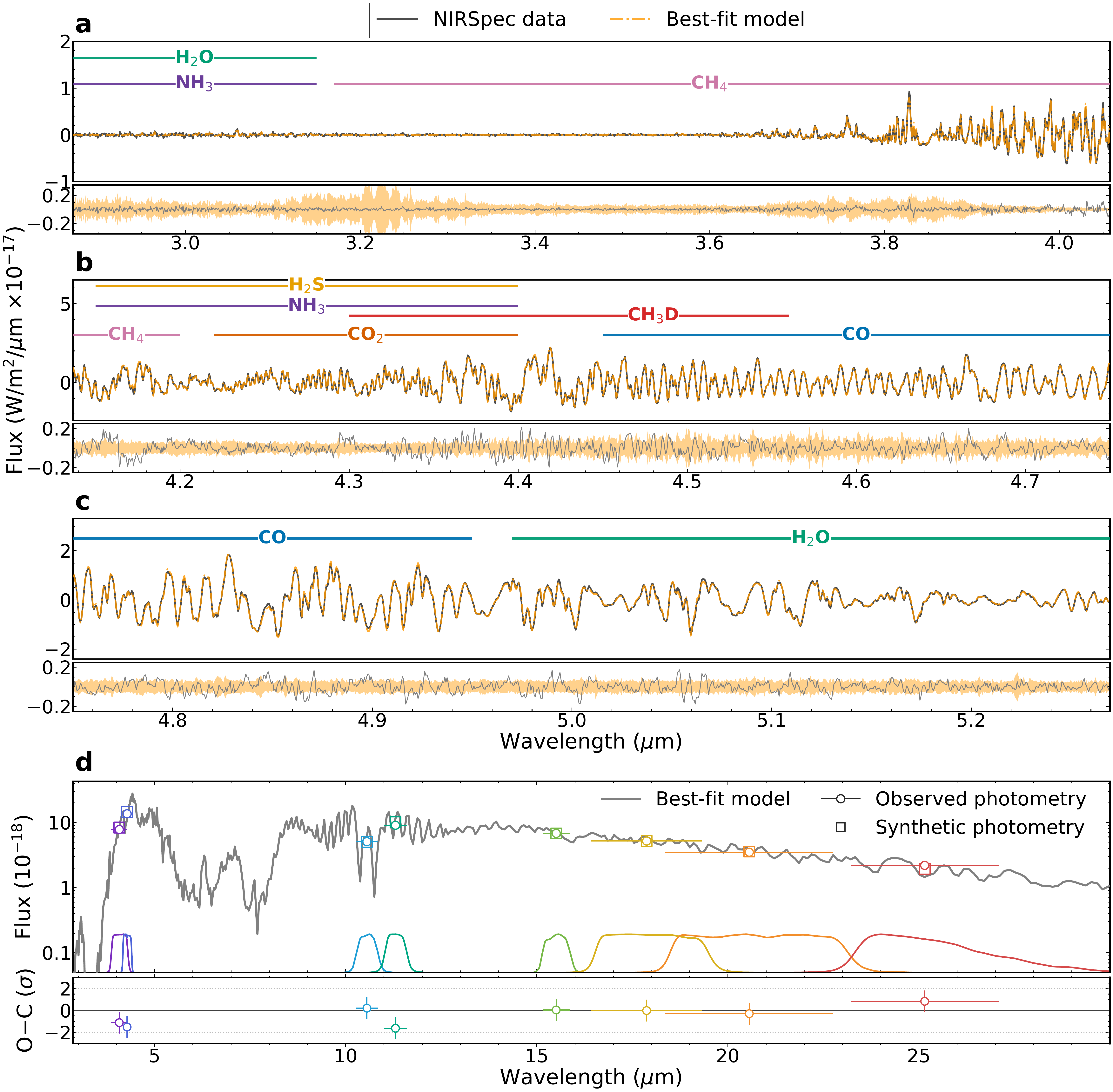}
    \caption{\textbf{JWST/NIRSpec spectrum and archival photometry of $\epsilon$ Ind Ab used in the atmospheric retrievals.} 
    \textbf{a,b,c,} NIRSpec spectrum from $2.8-5.3~\mu$m in black and best-fit model in orange. In each panel, we highlight the key absorbers for specific wavelength ranges. The bottom insets of each panel show the data residuals in gray, and inflated data uncertainties as orange contours.
    \textbf{d,} Circles show the JWST NIRCam and MIRI photometry covering $4-25~\mu$m \cite{Matthews2024, Matthews2026, Sanghi2026}. The gray curve is the best-fit model ($R=300$) made from correlated-k opacities, and open squares denote the synthetic photometry. The bottom inset shows data residuals in $\sigma$ units. All points are fit by the model well and agree at the $<2\sigma$ level.}
    \label{fig:nirspec_and_phot}
\end{figure}

\clearpage

\clearpage

\clearpage

\clearpage

\clearpage
\begin{figure}[h]
    \centering
    \includegraphics[width=0.55\linewidth]{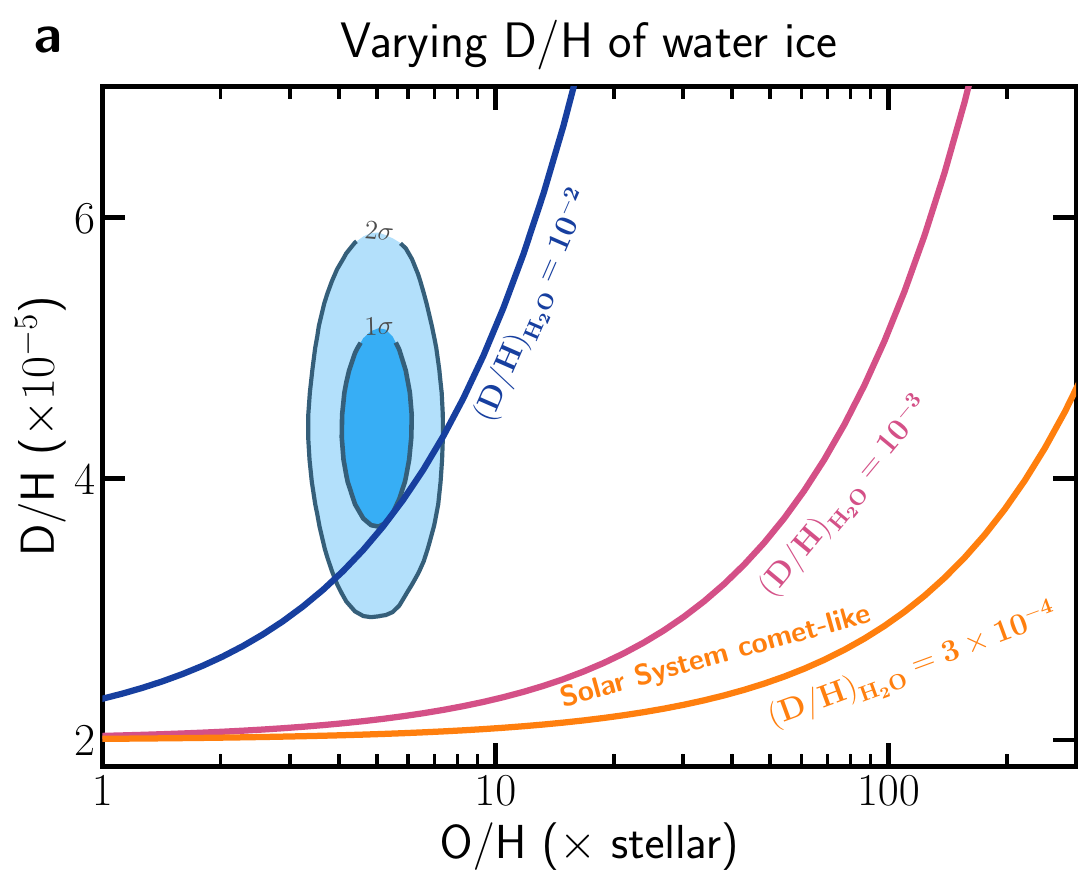}
    \includegraphics[width=0.55\linewidth]{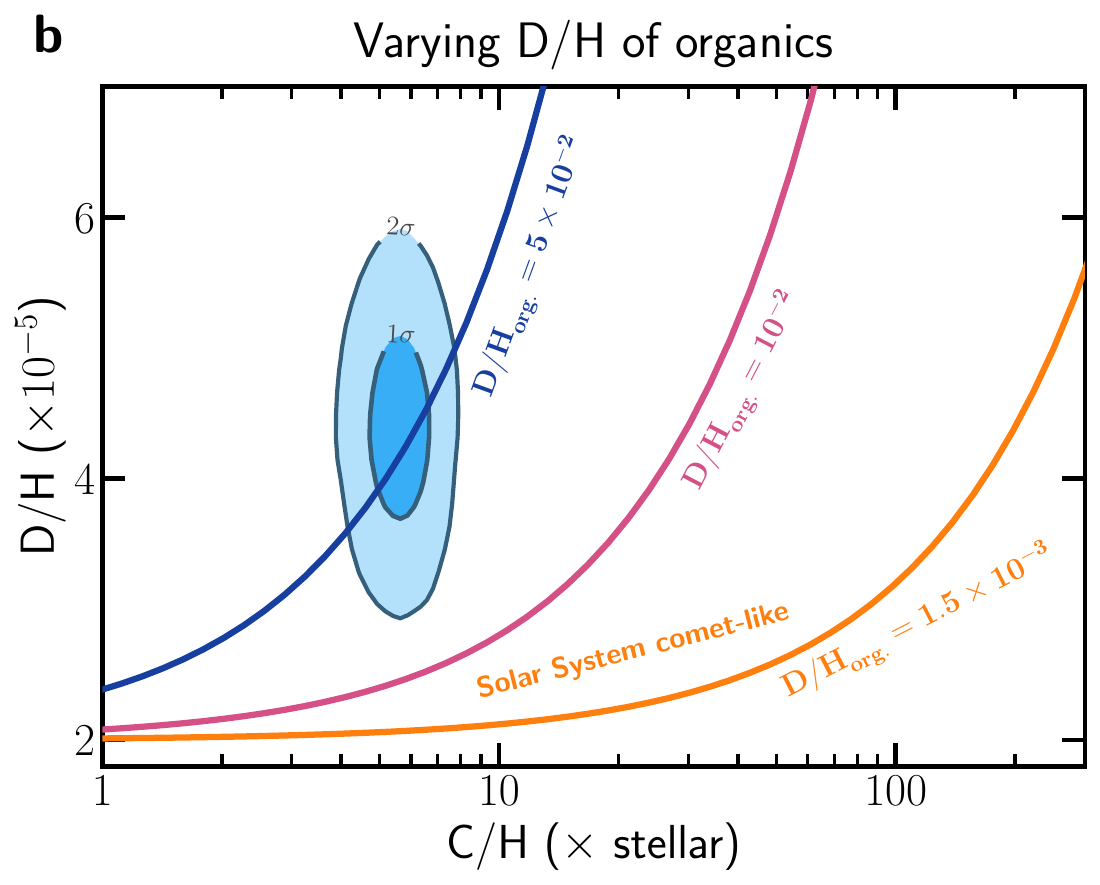}
    \caption{\textbf{Accretion of highly deuterium-rich solids is required to explain $\epsilon$ Ind Ab's D/H.} In the two panels, we plot the planetary bulk D/H as a function of the O and C enrichment. The colored lines correspond to different adopted D/H for water ice (\textbf{a}) and refractory organics (\textbf{b}). The D/H values adopted are chosen to represent objects in Fig.~\ref{fig:dtoh} (see references there). For water, $3 \times 10^{-4}$ represent a typical Solar System cometary value, $10^{-3}$ is a typical value for protostars, and $10^{-2}$ is the water D/H of 3I/ATLAS. For organics, $1.5\times10^{-3}$ is from comet 67P/C-G, $10^{-2}$ corresponds to CH$_2$DOH in protostars, and $5 \times 10^{-2}$ represents HCN in protoplanetary disks. If accreting material with Solar System comet-like D/H, matching the observed D/H requires very high enrichment in C/H and O/H of $\sim200 \times$ stellar, which is ruled out for $\epsilon$ Ind Ab. 
    }
    \label{fig:DtoH_vs_OtoH_CtoH}
\end{figure}

\clearpage

\begin{table*}
    \centering
    \caption{\textbf{Fitted parameters and priors for the atmospheric retrievals.}}
    \label{tab:param_prior}
    \small
    \begin{tabular}{ll|ll}
        \hline
        Parameter & Prior & Parameter & Prior \\
        \hline
        \multicolumn{4}{l}{\textit{Common parameters}} \\
        Mass ($\Mj$) & $\mathcal{N}(6.5,0.7)$ & NRS1 error multiple & $\mathcal{U}(1,10)$ \\
        Radius ($\Rj$) & $\mathcal{N}(1.04,0.04)$ & NRS2 error multiple & $\mathcal{U}(1,10)$ \\
        RV (\kms) & $\mathcal{U}(-60,-20)$ & $r_{\rm blue}$ & $\mathcal{U}(2400,3600)$ \\
        log(gray opacity) & $\mathcal{U}(-6,0)$ & $r_{\rm red}$ & $\mathcal{U}(3600,4600)$ \\
        \hline
        \multicolumn{4}{l}{\textit{P--T profile parameters}} \\
       $T_{\rm ref}$ [log($P$/bar)$=-1$] (K) & $\mathcal{U}(75,475)$ & $\nabla_7$ [0.0] & $\mathcal{N}(0.30,0.15)$ \\
       $\nabla_1$ [2.7] & $\mathcal{N}(0.26,0.06)$ & $\nabla_8$ [-1.0] & $\mathcal{N}(0.13,0.10)$ \\
        $\nabla_2$ [2.0] & $\mathcal{N}(0.28,0.06)$ & $\nabla_9$ [-2.0] & $\mathcal{N}(0.10,0.10)$ \\
        $\nabla_3$ [1.7] & $\mathcal{N}(0.29,0.06)$ & $\nabla_{10}$ [-3.0] & $\mathcal{N}(0.07,0.10)$ \\
        $\nabla_4$ [1.3] & $\mathcal{N}(0.29,0.06)$ & $\nabla_{11}$ [-4.0] & $\mathcal{U}(-0.40,0.50)$ \\
        $\nabla_5$ [1.0] & $\mathcal{N}(0.30,0.10)$ & $\nabla_{12}$ [-5.0] & $\mathcal{U}(-0.40,0.50)$ \\
        $\nabla_6$ [0.7] & $\mathcal{N}(0.30,0.10)$ & $\nabla_{13}$ [-6.0] & $\mathcal{U}(-0.40,0.50)$ \\
        $\nabla_7$ [0.0] & $\mathcal{N}(0.30,0.15)$ & & \\
        \hline
        \multicolumn{4}{l}{\textit{Free retrieval}} \\
        log mass-mixing ratio, $\log X_i$ & $\mathcal{U}(-12,-1)$ & & \\
        \hline
        \multicolumn{4}{l}{\textit{Chemically-consistent retrieval}} \\
        $[\mathrm{C/H}]$ & $\mathcal{U}(-0.5,1.0)$ & $\log(K_{\rm zz}/{\rm cm^2~s^{-1}})$ & $\mathcal{U}(2,9)$ \\
        $[\mathrm{O/H}]$ & $\mathcal{U}(-0.5,1.0)$ & log10 isotopologue ratio & $\mathcal{U}(1,10)$ \\
        $[\mathrm{N/H}]$ & $\mathcal{U}(-0.5,1.0)$ & & \\
        $[\mathrm{S/H}]$ & $\mathcal{U}(-0.5,1.0)$ & & \\
        \hline
    \end{tabular}
    \begin{flushleft}
    $\mathcal{U}$ denotes a uniform prior, and $\mathcal{N}$ denotes a Gaussian prior. 
    The P--T parameters $\nabla_i$ are the fitted $d\ln T/d\ln P$ values, with the associated pressure nodes given in square brackets in log(bar). 
    All logarithms are base 10. 
    In the free retrieval, each molecular species has an independent constant mass-mixing ratio. 
    In the chemically-consistent retrieval which uses \texttt{FastChem}, the molecular abundances are instead determined from the elemental abundances and $K_{\rm zz}$, while isotopologue ratios are fitted as log10 mass-fraction ratios relative to their parent isotopologues. 
    \end{flushleft}
\end{table*}

\pagebreak

\begin{table}
\centering
\caption{\textbf{Line opacities used in the atmospheric retrievals.}}
\label{tab:line_opacities}
\begin{tabular}{ll}
\hline
Species & Reference \\
\hline 
$^{12}$C$^{16}$O, $^{13}$C$^{16}$O, $^{12}$C$^{18}$O
    & \cite{Li2015} \\
H$_2^{16}$O
    & \cite{Polyansky2018} \\
H$_2^{18}$O, H$_2^{17}$O
    & \cite{Polyansky2017} \\
$^{12}$CH$_4$
    & \cite{Yurchenko2024} \\
$^{13}$CH$_4$, CH$_3$D
    & \cite{Hargreaves2020} \\
$^{14}$NH$_3$
    & \cite{Coles2019} \\
$^{12}$C$^{16}$O$_2$
    & \cite{Yurchenko2026} \\
H$_2$S
    & \cite{Azzam2016} \\
\hline
\end{tabular}
\end{table}

\begin{table}[htbp]
\centering
\caption{\textbf{Deuterium reactions included in \texttt{VULCAN}.}}
\label{tab:reactions}
\begingroup
\linespread{1}\selectfont   
\small                     
\setlength{\tabcolsep}{4pt}
\begin{tabular}{
    @{}r
    @{\hspace{4pt}} l
    @{\hspace{4pt}} r
    S[table-format=-1.3]
    S[table-format=-5.1]
    l@{}
}        
\toprule
No. & \quad Reaction$^{a}$
    & {$A$} & {$n$} & {$E_a$ (K)} & Ref. \\
\multicolumn{6}{l}{\quad Rate expression: $k=A\,T^n\exp(-E_a/T)$} \\
\midrule
\multicolumn{6}{l}{Bimolecular ($A$ in cm$^{3}$\,s$^{-1}$)} \\[2pt]
R1  & H$_2$ + D $\rightarrow$ H + HD
    & $3.98\times10^{-18}$ & 2.290 &  2630.0 & \cite{Michael1990_MIC_FIS} \\
R3  & D$_2$ + O $\rightarrow$ OD + D
    & $1.99\times10^{-16}$ & 1.720 &  4880.0 & \cite{Yang1993} \\
R5  & D$_2$ + OD $\rightarrow$ D$_2$O + D
    & $1.60\times10^{-17}$ & 1.870 &  1724.0 & \cite{Nguyen2011} \\
R7  & D$_2$ + H $\rightarrow$ HD + D
    & $1.69\times10^{-17}$ & 2.100 &  3530.0 & \cite{Michael1990b} \\
R9  & D$_2$ + OH $\rightarrow$ HDO + D
    & $3.30\times10^{-17}$ & 1.770 &  1830.0 & \cite{Nguyen2011} \\
R11 & O($^1$D) + D$_2$ $\rightarrow$ OD + D
    & $1.10\times10^{-10}$ & 0.000 &     0.0 & \cite{Talukdar1996} \\
R13 & OH + HD $\rightarrow$ HDO + H
    & $7.87\times10^{-12}$ & 0.000 &  2810.0 & \cite{Zhang1995} \\
R15 & OH + HD $\rightarrow$ H$_2$O + D
    & $4.70\times10^{-12}$ & 0.000 &  2260.0 & \cite{Zhang1995} \\
R17 & OH + D $\rightarrow$ OD + H
    & $1.90\times10^{-9}$  & -0.630 &    0.0 & \cite{Howard1982} \\
R19 & H$_2$ + OD $\rightarrow$ H + HDO
    & $2.80\times10^{-17}$ & 1.830 &  1419.0 & \cite{Nguyen2011} \\
R21 & CH$_4$ + OD $\rightarrow$ CH$_3$ + HDO
    & $3.90\times10^{-19}$ & 2.390 &  1270.0 & \cite{Masgrau2001} \\
R23 & CH$_3$D + OH $\rightarrow$ CH$_2$D + H$_2$O
    & $5.19\times10^{-18}$ & 2.000 &  1332.0 & \cite{Atkinson2003} \\
R25 & CH$_3$D + OH $\rightarrow$ CH$_3$ + HDO
    & $6.70\times10^{-19}$ & 2.320 &  1430.0 & \cite{Masgrau2001} \\
R27 & CH$_3$ + D $\rightarrow$ CH$_2$D + H
    & $2.31\times10^{-10}$ & 0.000 &     0.0 & \cite{Seakins1997} \\
R29 & CH$_3$ + D $\rightarrow$ CH$_3$D
    & $2.74\times10^{-10}$ & 0.000 &   106.0 & \cite{Hase1987} \\
R31 & CH$_3$ + HD $\rightarrow$ CH$_3$D + H
    & $2.23\times10^{-20}$ & 2.514 &  4295.0 & \cite{FernndezRamos2001} \\
R33 & CH$_4$ + D $\rightarrow$ CH$_3$ + HD
    & $7.47\times10^{-11}$ & 0.000 &  5590.0 & \cite{Kurylo1970} \\
R35 & H$_2$O + OD $\rightarrow$ HDO + OH
    & $7.23\times10^{-16}$ & 1.980 &  -527.0 & \cite{Masgrau1999} \\
R37 & H + CH$_3$D $\rightarrow$ CH$_2$D + H$_2$
    & $1.65\times10^{-20}$ & 3.000 &  4040.0 & assumed $\tfrac{3}{4}\times$ H + CH$_4$ \\
R39 & CH$_3$ + D$_2$ $\rightarrow$ CH$_3$D + D
    & $2.36\times10^{-11}$ & 0.000 &  7810.0 & \cite{Rodriguez1986} \\
\midrule
\multicolumn{6}{l}{Termolecular (first row: low-pressure limit $k_0$ in cm$^{6}$\,s$^{-1}$;} \\
\multicolumn{6}{l}{\phantom{Termolecular (}second row: high-pressure limit $k_\infty$ in cm$^{3}$\,s$^{-1}$)} \\[2pt]
R41 & D + D + M $\rightarrow$ D$_2$ + M
    & $6.09\times10^{-33}$ &  0.000 &     0.0 & \cite{Trainor1973, Takahashi1977} \\
    &
    & $5.00\times10^{-16}$ &  0.000 &     0.0 & \cite{Takahashi1977} \\
R43 & D + H + M $\rightarrow$ HD + M
    & $2.70\times10^{-31}$ & -0.600 &     0.0 & assumed same as H + H $\rightarrow$ H$_2$ \\
    &
    & $3.31\times10^{-6}$  & -1.000 &     0.0 & \\[2pt]
R45 & CH$_2$D + H + M $\rightarrow$ CH$_3$ + D + M
    & $1.99\times10^{-31}$ & -1.090 &  1100.0 & \cite{Stewart1989} \\
    &
    & $5.15\times10^{-22}$ &  2.300 &   207.0 & \\[2pt]
R47 & CH$_3$D + M $\rightarrow$ CH$_2$D + H + M
    & $1.16\times10^{12}$  & -4.480 & 53800.0 & \cite{Stewart1989} \\
    &
    & $5.57\times10^{16}$  & -0.320 & 52300.0 & \\[2pt]
R49 & CH$_3$D + M $\rightarrow$ CH$_3$ + D + M
    & $9.69\times10^{8}$   & -3.850 & 54700.0 & \cite{Stewart1989} \\
    &
    & $1.49\times10^{18}$  & -0.890 & 53800.0 & \\
\midrule
\multicolumn{6}{l}{Photodissociation} \\[2pt]
R51 & \multicolumn{4}{l}{$\mathrm{HDO \rightarrow H + OD}$}
    & \makecell[l]{cross-sections from \cite{Chung2001,Cheng2004};\\[-2pt]
                   branching ratios assumed\\[-2pt]
                   the same as H$_2$O} \\
R53 & \multicolumn{4}{l}{$\mathrm{HDO \rightarrow D + OH}$} & \\
R55 & \multicolumn{4}{l}{$\mathrm{CH_3D \rightarrow CH_2D + H}$}
    & assumed same as CH$_4$ \\
R57 & \multicolumn{4}{l}{$\mathrm{CH_3D \rightarrow CH_3 + D}$} & \\
R59 & \multicolumn{4}{l}{$\mathrm{CH_3D \rightarrow CH_2 + HD}$} & \\
\bottomrule
\multicolumn{6}{l}{\footnotesize $^{a}$ Even-numbered reactions are backward reactions reversed by thermodynamic data.} \\
\end{tabular}
\endgroup
\end{table}


\clearpage 


\end{document}